\documentclass[prd,twocolumn,showpacs,superscriptaddress,nofootinbib,floatfix,showkeys,10pt]{revtex4-1}
\usepackage{graphicx}
\usepackage{bm}
\usepackage{yhmath}
\usepackage{mathtools}
\usepackage{wasysym}
\usepackage[colorlinks,citecolor=violet,urlcolor=violet,linkcolor=blue]{hyperref}
\usepackage{color}
\usepackage{cases}
\usepackage{times}
\usepackage{dcolumn,booktabs,bm}
\usepackage{slashed}
\usepackage{amsfonts,amssymb,stmaryrd,latexsym,amsmath}
\usepackage{textcomp}
\usepackage{cancel}
\usepackage{array}
\usepackage{orcidlink}
\usepackage{physics}
\usepackage{xfrac}
\usepackage{multirow}
\usepackage{subfig}

\allowdisplaybreaks

\begin{document}
\title{Constraints on the mass of the dark antibaryon using $B_d\rightarrow \Lambda  \psi_{DS}$ channel in light cone QCD}
\author{M. A. Abri\,\orcidlink{0009-0003-8050-2413}}
\affiliation{Department of Physics, University of Tehran, North Karegar Avenue, Tehran
    14395-547, Iran}
\author{N. Hajirasouliha\,\orcidlink{0009-0004-6130-6844}}
\affiliation{Department of Physics, University of Tehran, North Karegar Avenue, Tehran
    14395-547, Iran}
\author{K. Azizi\,\orcidlink{0000-0003-3741-2167}}\thanks{Corresponding Author}\email{kazem.azizi@ut.ac.ir}
\affiliation{Department of Physics, University of Tehran, North Karegar Avenue, Tehran 14395-547, Iran}
\affiliation{Department of Physics, Dogus University, Dudullu-\"{U}mraniye, 34775 Istanbul, T\"urkiye}
%%%%%%%%%%%%%%%%%%%%%%%%%%%%%%%%%%%%%%%%%%%%%%%%%%%%%%%%%%%%%%%%%
\begin{abstract}
    According to the $B$-mesogenesis framework, the baryon asymmetry of the Universe and dark matter can be simultaneously generated through CP-violating $B$-meson oscillations. In this mechanism, $B$-mesons decay into a Standard Model baryon and a dark-sector antibaryon, denoted by $\psi_{DS}$. Within this scenario, we investigate the allowed mass window for $\psi_{DS}$ using Light Cone Sum Rules (LCSR) for $B_d\rightarrow\Lambda \, \psi_{DS}$ decay. To include nonperturbative effects, we employ contributions up to twist-6 of the $\Lambda$ distribution amplitudes in the operator product expansion (OPE). We derive the branching fraction as a function of dark antibaryon mass and, by comparing with the experimental limits by the BaBar and Belle collaborations, determine the mass ranges of $\psi_{DS}$ consistent with the $B$-mesogenesis mechanism.
\end{abstract}
\maketitle
%%%%%%%%%%%%%%%%%%%%%%%%%%%%%%%%%%%%%%%%%%%%%%%%%%%%%%%%%%%%%%%%%
\section{Introduction}
\label{sec:Introduction}
The Standard Model (SM) of particle physics has been highly successful in describing particles and their interactions. However, it has failed to address certain well-known problems, indicating the need for physics beyond the Standard Model (BSM). Among these are the nature of dark matter (DM) and the reason behind the baryon asymmetry of the universe (BAU) \cite{Planck:2015fie, Planck:2018vyg, Cyburt:2015mya, Mossa:2020gjc}. Although baryogenesis satisfies Sakharov conditions \cite{Sakharov:1967dj}, it is quantitatively insufficient \cite{Arnold:1992rz, Cohen:1993nk, Gavela:1993ts, Kirilova:1995rf, Kajantie:1996mn, Losada:1996rt, Trodden:1998qg, Riotto:1999yt, Dine:2003ax, Morrissey:2012db, DOnofrio:2015gop, vandeVis:2025efm}. In principle, conventional baryogenesis and leptogenesis mechanisms meet the required conditions but they encounter theoretical or experimental challenges \cite{Dimopoulos:1978kv, Fukugita:1986hr, Affleck:1984fy, Buchmuller:2002rq, Pilaftsis:2003gt, Giudice:2003jh, Buchmuller:2005eh, Nardi:2006fx, Davidson:2008bu}.

There are two possible explanations for the observed matter-antimatter asymmetry in the universe. One is that the laws of physics are symmetric but the initial conditions are not, meaning they are slightly matter-dominated \cite{Rafelski:2023emw, Pereira:2023xiw}. This looks simple enough but it merely shifts the question to why the initial conditions are asymmetric. The second possibility is that the laws of physics themselves violate the symmetry between matter and antimatter. We already know that in this case, the SM does not respect CP symmetry, but doesn't break it enough to account for the BAU. Depending on when the asymmetry is first generated, different BSM scenarios arise. If a baryon asymmetry is generated directly, the scenario is referred to as baryogenesis. If a lepton asymmetry is generated at first and then converted into a baryon asymmetry via electroweak sphaleron processes, the mechanism is referred to as leptogenesis \cite{Heisig:2024mwr, Blazek:2025wmc, Falkowski:2017uya}. In the scenario considered in this paper, the SM itself could suffice if meson decays produce both baryons and dark antibaryons, which is called Mesogenesis \cite{Ghalsasi:2015mxa, McKeen:2015cuz, Elor:2018twp, Nelson:2019fln, Alonso-Alvarez:2019fym, Elor:2020tkc, Elahi:2021jia, Alonso-Alvarez:2021oaj, Elor:2024cea, Elor:2025fcp, Miro:2025mck, BurgosMarcos:2026lwv}.

As a low-energy and late-time baryogenesis mechanism, mesogenesis relates BAU to different decays of SM mesons. These mesons include $\{D^{\pm},\ B^{\pm},\ B^0_{d,s},\ B^{\pm}_c\}$. In these scenarios, BAU is directly proportional to the CP violation in meson mixing. In the model proposed for $B$-meson known as $B$-mesogenesis, a long-lived scalar field in the pre-nucleosynthesis universe decays into b-quarks and antiquarks. The temperature should be low enough that hadronization can occur. Subsequently, through $B-\bar B$ CP-violating oscillations and their hadronic decay to dark sector, the BAU becomes the asymmetry of the visible sector. In recent years, there has been evidence of relatively large CP violation in $B$-mesons \cite{LHCb:2025ray, Belle:2001zzw, Fleischer:2024uru, Miro:2024fid} which can generate the required BAU \cite{Elor:2018twp, Elor:2024cea, Elor:2025fcp}. Consequently, precision measurements for branching fractions of rare decay modes provide an experimental probe of $B$-mesogenesis.

In recent years, considerable theoretical effort has been devoted to searches for rare and invisible decays of $B$ mesons into dark-sector states. The decay $B^+\rightarrow p \psi_{DS}$ was investigated at leading twist using the LCSR approach in Ref.~\cite{Khodjamirian:2022vta}, and its higher-twist corrections were studied in Ref.~\cite{Boushmelev:2023huu} using proton distribution amplitudes (DAs). The same decay channel was also analyzed within LCSR employing $B$-meson DAs instead, as reported in Ref.~\cite{Biswas:2026oxq}. Various two-body decays of the $B$-meson set $\{B^+,B_d,B_s\}$ into octet and charmed anti-triplet baryons were investigated using LCSR with leading twist-3 baryon DAs in Ref.~\cite{Elor:2022jxy}. Semi-inclusive decays $B\rightarrow X_{u/c,d/s}\psi_{DS}$ were studied within the heavy-quark expansion (HQE) framework in Ref.~\cite{Shi:2023riy}, where $X_{u/c,d/s}$ denotes hadrons containing $u/c$ and $d/s$ quarks with unit baryon number. The fully inclusive decay rate $\Gamma(b\rightarrow du\psi_{DS})$ was computed within the HQE in Ref.~\cite{Lenz:2024rwi}, where lower bounds on the $b\rightarrow X \psi_{DS}$ couplings were also estimated. In a more recent study, the inclusive rate was calculated up to the dimension-six two-quark Darwin operator, and regions of parameter space in which subleading terms exceed the leading contribution were identified \cite{Mohamed:2025zgx}. Additionally, bottom-baryon decays into light mesons have been investigated using various approaches in Refs.~\cite{Zheng:2024tkj,Xing:2025pfw,Shi:2024uqs}.

In this work, we employ the powerful method of LCSR to calculate the branching fraction of $B_d\rightarrow \Lambda\ \psi_{DS}$(we will use $\psi$ for dark antibaryon state throughout the paper). In the $B$-mesogenesis framework, this channel is mediated by a heavy colored scalar field with a mass at the TeV scale. Two operators are defined for the effective Lagrangian based on the pairing of the spinors. For the $B$-meson state, we use its standard interpolating current, while for the $\Lambda$ baryon, we employ octet DAs up to twist-6, related to those of the $\Lambda$ through $SU(3)_f$ symmetry within the conformal partial wave expansion approach. The $SU(3)_f$ symmetry‑breaking effects specific to the $\Lambda$ baryon have been incorporated following Ref.~\cite{Liu:2008yg}, and isospin‑symmetry breaking as well as corrections to the nonperturbative parameters have been estimated for all fourteen independent distribution amplitudes. The resulting form factors of this transition are extrapolated using $z$-expansion to the physical region. On the experimental front, Belle, BaBar and LHCb collaborations have investigated the exotic decay channels of $b$-quark \cite{BaBar:2024qqx,Rodriguez:2021urv,Belle:2021gmc,BaBar:2023rer,Ahmed:2024tui}. From upper limits of the branching fraction corresponding to this channel reported by Belle and BaBar, we derive constraints on the allowed mass range of $\psi$.

The paper is organized as follows. Section~\ref{sec:Effective_inter} presents the effective interaction governing the decay. The theoretical framework based on the LCSR approach is established by constructing the correlation function in Sec.~\ref{sec:LCSR}. The hadronic (physical) and QCD representations of the corresponding LCSRs for the two models considered are presented in Subsecs.~\ref{subsec:physical_side} and \ref{subsec:QCD_side}, respectively. The numerical analysis of the resulting sum rules is presented in Sec.~\ref{sec:Numerical_analysis}. The form factors are extrapolated in Subsec.~\ref{subsec:FFs}, and the branching fractions obtained from these form factors are compared with the experimental limits in Subsec.~\ref{subsec:Br}. Finally, Sec.~\ref{sec:conclusion} summarizes our results and presents the concluding remarks.
%%%%%%%%%%%%%%%%%%%%%%%%%%%%%%%%%%%%%%%%%%%%%%%%%%%%%%%%%%%%%%%%%%%

\section{Effective interaction}
\label{sec:Effective_inter}
Within the $B$-mesogenesis framework, the renormalizable Lagrangian is determined by the hypercharge assignment of the mediator field $Y$, which is a color-triplet scalar field with mass $M_Y$ \cite{Alonso-Alvarez:2021qfd,Elor:2018twp,Elor:2022jxy}. In this work, we consider the scenario with hypercharge $-1/3$. For the UV completion of the model, see Ref. \cite{Alonso-Alvarez:2019fym}. The decay of a $\bar b$-quark inside a $B_d$ meson to a $\Lambda$ baryon and dark-sector antibaryon $\psi$ is represented by the effective Lagrangian
\begin{equation}
    \begin{aligned}
        \mathcal{L}_{(-1/3)} & =-y_{ub}\ \epsilon_{ijk}\ Y^{*i}\bar{u}_R^j\ b_R^{c,k}-y_{us}\ \epsilon_{ijk}\ Y^{*i}\ \bar{u}_R^j s_R^{c,k} \\&-y_{\psi b}Y_i\ \bar{\psi}\ b_R^{c,i}-y_{\psi s}\ Y_i\ \bar{\psi}\ s_R^{c,i}+\text{h.c.},
    \end{aligned}
    \label{eq:Lagrangian}
\end{equation}
where $q_R^c$ denotes the charge-conjugated right-handed quark fields, $\psi$ is the dark antibaryon field (observable as missing energy in detectors), and $i, j, k$ are the color indices in the fundamental representation. The dimensionless constants $y$ denote the relevant couplings. The terms proportional to $y_{ub}$ and $y_{us}$ involve antisymmetric color contractions, ensuring invariance under $SU(3)_c$. In contrast, the terms with $y_{\psi b}$ and $y_{\psi s}$ are color-singlets and represents interactions linking SM to the dark sector.

Since $M_Y$ is significantly larger than the momentum transfer of the decay $k\sim m_{B_d}$, the heavy mediator can be integrated out, reducing its propagator to:
\begin{equation}
    \begin{aligned}
        \frac{i}{k^2-M_{Y}^2}=-\frac{i}{M_{Y}^2}\left(1+\frac{k^2}{M_{Y}^2}+...\right)\approx -\frac{i}{M_{Y}^2}.
    \end{aligned}
    \label{eq:Y_propagator}
\end{equation}
Applying this substitution to Eq.~\eqref{eq:Lagrangian} yields a low-energy effective Lagrangian containing interactions between $\psi$ and three quarks. To extract three-quark operators for the hadronic decay, it is convenient to define two models depending on which quark couples to the dark state $\psi$. Following Ref. \cite{Alonso-Alvarez:2021qfd}, type-I and II operators can be defined corresponding to the $(\boldsymbol b)$ and $(\boldsymbol s)$-model scenarios, respectively: 
\begin{subequations}
    \begin{equation}
     \mathcal{L}^{(\boldsymbol s)}_{\text{eff}} = i\epsilon_{ijk}\frac{y_{ub}y_{\psi s}}{M^2_Y}(\bar{\psi}s_R^{c,i})(\bar{u}_R^j b_R^{c,k})+\text{h.c.},       
     \label{eq:L_eff_s}
    \end{equation}
    \begin{equation}
                \mathcal{L}^{(\boldsymbol b)}_{\text{eff}} = i\epsilon_{ijk}\frac{y_{us}y_{\psi b}}{M^2_Y}(\bar{\psi}b_R^{c,i})(\bar{u}_R^j s_R^{c,k})+\text{h.c.}, 
        \label{eq:L_eff_b}
    \end{equation}
\end{subequations}
 from which we can verify the interacting Hamiltonians as:
\begin{subequations}
    \begin{equation}
            \mathcal{H}^{(\boldsymbol s)}_{(-1/3)}  =  -i\epsilon_{ijk}\frac{y_{ub}y_{\psi s}}{M^2_Y}(\bar{\psi}s_R^{c,i})(\bar{u}_R^j b_R^{c,k})+\text{h.c.},        
        \label{eq:H_eff_s}
    \end{equation}
    \begin{equation}    
            \mathcal{H}^{(\boldsymbol b)}_{(-1/3)}  = -i\epsilon_{ijk}\frac{y_{us}y_{\psi b}}{M^2_Y}(\bar{\psi}b_R^{c,i})(\bar{u}_R^j s_R^{c,k})+\text{h.c.}~. 
        \label{eq:H_eff_b}
    \end{equation}
\end{subequations}
Using generalized Fierz identities we can factorize the dark antibaryon field $\psi$ \cite{Nieves:2003in}:
\begin{equation}
    \begin{aligned}
        \bar{\psi}\, s_R^c & = \bar{s}_R \, \psi^c, & \quad \bar{s}_R^c \, \psi & = \bar{\psi}^c \, s_R, \\
        \bar{\psi}\, b_R^c & = \bar{b}_R \, \psi^c, & \quad \bar{b}_R^c \, \psi & = \bar{\psi}^c \, b_R.
    \end{aligned}
\end{equation}
Applying these identities to Eq.s ~\eqref{eq:H_eff_s} and ~\eqref{eq:H_eff_b},  we obtain:
\begin{subequations}
    \begin{equation}
        \begin{aligned}
            \mathcal{H}^{(\boldsymbol s)}_{(-1/3)}=-G^{(\boldsymbol s)} \bar{\mathcal{O}}^{(\boldsymbol s)} \psi^c -\text{h.c.},
        \end{aligned}
        \label{eq:H_eff_s_final}
    \end{equation}
    \begin{equation}
        \begin{aligned}
            \mathcal{H}^{(\boldsymbol b)}_{(-1/3)}=-G^{(\boldsymbol b)} \bar{\mathcal{O}}^{(\boldsymbol b)} \psi^c -\text{h.c.},
        \end{aligned}
        \label{eq:H_eff_b_final}
    \end{equation}
\end{subequations}
where
\begin{equation}
        G^{(\boldsymbol s)}  = \frac{y_{ub}y_{\psi s}}{M^2_Y},\quad G^{(\boldsymbol b)}  = \frac{y_{us}y_{\psi b}}{M^2_Y},
    \label{eq:eff_couplings}
\end{equation}
are the effective four-fermion couplings, and
\begin{subequations}
    \begin{equation}
            \mathcal{O}^{(\boldsymbol s)} = i\epsilon_{ijk}s_R^i(\bar{b}_R^{c,j} u_R^k),
        \label{eq:O_s}
    \end{equation}
    \begin{equation}
            \mathcal{O}^{(\boldsymbol b)} = i\epsilon_{ijk}b_R^i(\bar{s}_R^{c,j} u_R^k),
        \label{eq:O_b}
    \end{equation}
\end{subequations}
are the effective local three-quark operators. The decay amplitude of each model are given by
\begin{subequations}
    \begin{equation}
        \begin{aligned}
            \mathcal{A}^{(\boldsymbol s)}(B_d  \to  \Lambda  \psi)  &=  G^{(\boldsymbol s)}  \langle \Lambda(P)\psi^c(q) | \bar{\mathcal{O}}^{(\boldsymbol s)} | B_d(P+q)\rangle \\ &= G^{(\boldsymbol s)}  \langle \Lambda(P)| \bar{\mathcal{O}}^{(\boldsymbol s)} | B_d(P+q)\rangle u_\psi^c(q),
        \end{aligned}
        \label{eq:Amplitude_s}
    \end{equation}
    \begin{equation}
        \begin{aligned}
            \mathcal{A}^{(\boldsymbol b)}(B_d  \to \Lambda \psi) & = G^{(\boldsymbol b)}  \langle \Lambda(P)\psi^c(q) | \bar{\mathcal{O}}^{(\boldsymbol b)} | B_d(P+q)\rangle \\ &= G^{(\boldsymbol b)}  \langle \Lambda(P)| \bar{\mathcal{O}}^{(\boldsymbol b)} | B_d(P+q)\rangle u_\psi^c(q),
        \end{aligned}
        \label{eq:Amplitude_b}
    \end{equation}
\end{subequations}
where $u_\psi^c$ is the charge conjugated Dirac bispinor of the dark antibaryon field $\psi$. The four-momentum of the involved particles are specified in Eq.~\eqref{eq:Amplitude_s} and ~\eqref{eq:Amplitude_b}, where $(P+q)^2=m_{B_d}^2$, $P^2=m_\Lambda^2$ and $q^2=m_\psi^2$. The matrix elements involving the operators $\mathcal{O}^{(\boldsymbol s),(\boldsymbol b)}$ yield zero. The hadronic matrix elements can be decomposed for each model as a linear independent combination of four transition form factors. For instance, for the $(\boldsymbol s)$-model \cite{Khodjamirian:2022vta,Boushmelev:2023huu}:
\begin{equation}
\begin{aligned}
\langle \Lambda(P) | \bar{\mathcal O}^{(\boldsymbol s)} | B_d(P+q) \rangle
&= F^{(\boldsymbol s)}_{B_d\to\Lambda_R}(q^2)\bar u_{\Lambda_R}(P)
\\& + F^{(\boldsymbol s)}_{B_d\to\Lambda_L}(q^2)\bar u_{\Lambda_L}(P)
\\& + \tilde F^{(\boldsymbol s)}_{B_d\to\Lambda_R}(q^2)\bar u_{\Lambda_R}(P)\frac{\slashed q}{m_\Lambda}
\\& + \tilde F^{(\boldsymbol s)}_{B_d\to\Lambda_L}(q^2)\bar u_{\Lambda_L}(P)\frac{\slashed q}{m_\Lambda},
\end{aligned}
\label{eq:FormFactors_s}
\end{equation}
where $u_{\Lambda_{R,L}}=\frac{1}{2}(1\pm \gamma_5)u_\Lambda$ are the right- and left-handed components of the $\Lambda$-baryon spinor, and the form factors are written as four independent kinematical structures considering the four-momenta of the involved particles. Since the operator $\mathcal O^{(\boldsymbol s)}$ possesses a definite chirality,
satisfying $\bar{\mathcal O}^{(\boldsymbol s)}P_L =\bar{\mathcal O}^{(\boldsymbol s)}$, the two form factors $F_{B_d\to\Lambda_L}^{(\boldsymbol s)}$ and $\tilde F_{B_d\to\Lambda_R}^{(\boldsymbol s)}$ vanish identically. The same decomposition can be done for the $(\boldsymbol b)$-model, where an analogous chirality argument eliminates the corresponding two form factors $F_{B_d\to\Lambda_L}^{(\boldsymbol b)}$ and $\tilde F_{B_d\to\Lambda_R}^{(\boldsymbol b)}$. From this point on, we restrict our calculations to the $(\boldsymbol s)$-model, since the procedure is identical for both models.

At fixed $q^2=m_\psi^2$ and considering the Dirac equation, the $(\boldsymbol s)$-model decay amplitude can be rewritten as:
\begin{equation}
    \begin{aligned}
          \mathcal{A}^{(\boldsymbol s)}\left(B_d \rightarrow \Lambda\psi\right)&= G^{(\boldsymbol s)}\bigg[F^{(\boldsymbol s)}_{B_d\rightarrow\Lambda_R}(q^2)\bar{u}_\Lambda(P) P_L
            \\&+ F^{(\boldsymbol s)}_{B_d\rightarrow\Lambda_L}(q^2)\bar{u}_\Lambda(P) P_R
       \\  & +\frac{m_\psi}{m_\Lambda}\tilde{F}^{(\boldsymbol s)}_{B_d\rightarrow\Lambda_R}(q^2)\bar{u}_\Lambda(P) P_L
        \\ &+                                 \frac{m_\psi}{m_\Lambda}\tilde{F}^{(\boldsymbol s)}_{B_d\rightarrow\Lambda_L}(q^2)\bar{u}_\Lambda(P) P_R\bigg] u_\psi^c(q),
    \end{aligned}
    \label{eq:DecayAmp_proj}
\end{equation}
where $P_{R,L}=\frac{1}{2}(1\pm \gamma_5)$ are the left and right-handed projection operators. The decay amplitude can then be simplified as:
\begin{equation}
        \mathcal{A}^{(\boldsymbol s)}\left(B_d \rightarrow \Lambda\psi\right) =
        G^{(\boldsymbol s)}\bar{u}_\Lambda(P)\big[ A^{(\boldsymbol s)} + \gamma_5 B^{(\boldsymbol s)} \,\big]\,u_\psi^c(q),
    \label{eq:DecayAmp_simplified}
\end{equation}
with
\begin{equation}
    \begin{aligned}
        A^{(\boldsymbol s)}  = \tfrac{1}{2}  & \Big[ F^{(\boldsymbol s)}_{B_d\rightarrow\Lambda_R}(q^2) + \tilde{F}^{(\boldsymbol s)}_{B_d\rightarrow\Lambda_R}(q^2)\tfrac{m_\psi}{m_\Lambda}
        \\ +& F^{(\boldsymbol s)}_{B_d\rightarrow\Lambda_L}(q^2) + \tilde{F}^{(\boldsymbol s)}_{B_d\rightarrow\Lambda_L}(q^2)\tfrac{m_\psi}{m_\Lambda} \Big],                 \\[12pt]
        B^{(\boldsymbol s)}                  = \tfrac{1}{2} & \Big[ -F^{(\boldsymbol s)}_{B_d\rightarrow\Lambda_R}(q^2) - \tilde{F}^{(\boldsymbol s)}_{B_d\rightarrow\Lambda_R}(q^2)\tfrac{m_\psi}{m_\Lambda}
        \\  +& F^{(\boldsymbol s)}_{B_d\rightarrow\Lambda_L}(q^2) + \tilde{F}^{(\boldsymbol s)}_{B_d\rightarrow\Lambda_L}(q^2)\tfrac{m_\psi}{m_\Lambda} \Big].
    \end{aligned}
    \label{eq:AB_definitions}
\end{equation}
The two-body decay width, expressed in terms of  the two functions defined in Eq.~\eqref{eq:AB_definitions}, is given by
\begin{equation}
\begin{aligned}
\Gamma^{(\boldsymbol s)}(B_d \to \Lambda \psi)
&= \frac{|G^{(\boldsymbol s)}|^2}{8\pi m_{B_d}^3}\Bigl\{|A^{(\boldsymbol s)}|^2\bigl[m_{B_d}^2 - (m_\Lambda - m_\psi)^2\bigr]
\\&+ |B^{(\boldsymbol s)}|^2\bigl[m_{B_d}^2 - (m_\Lambda + m_\psi)^2\bigr]\Bigr\}
\\&\times\lambda^{1/2}(m_{B_d}^2,m_\psi^2,m_\Lambda^2),
\end{aligned}
\label{eq:two_body_decay_final}
\end{equation}
where $\lambda$ is the Källen function. The same relations are obtained for $(\boldsymbol b)$-model by replacing $\boldsymbol s\to \boldsymbol b$ in Eqs.~(\ref{eq:FormFactors_s}--\ref{eq:two_body_decay_final}).

%%%%%%%%%%%%%%%%%%%%%%%%%%%%%%%%%%%%%%%%%%%%%%%%%%%%%%%%%%%

\section{FORMALISM}
\label{sec:LCSR}
\begin{figure}[tb]
  \centering
  \includegraphics[width=1\linewidth]{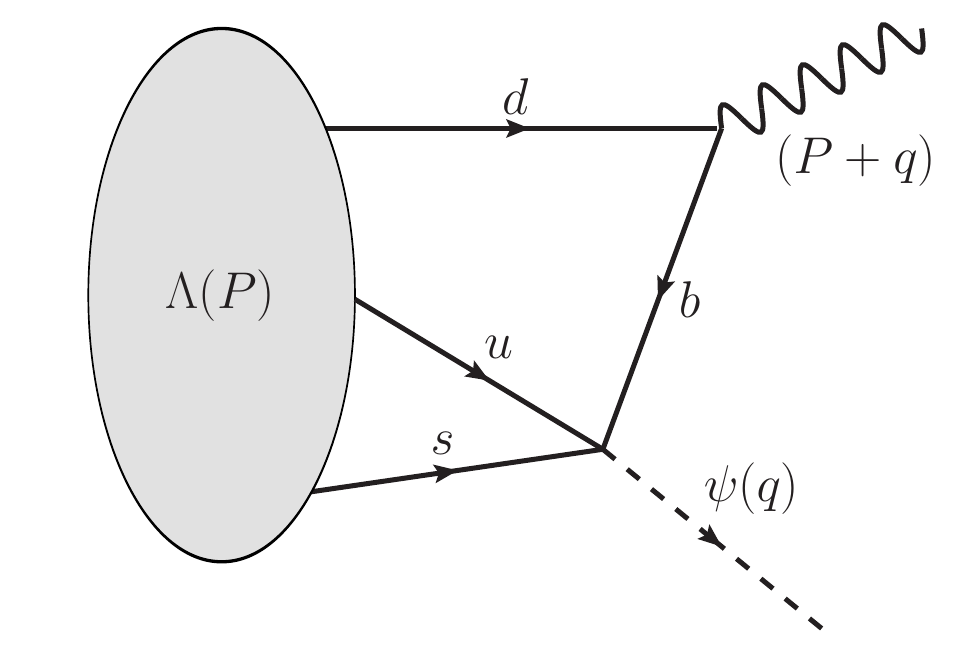}
 \caption{The diagram corresponding to the both ($\boldsymbol s$)- and ($\boldsymbol b$)- models.}
  \label{fig:leadingDiagram}
\end{figure}

To study the form factors in each model, we define the two-point correlation function corresponding to the $B_d\rightarrow \Lambda\psi$ decay mode. Considering the $(\boldsymbol s)$-model, the correlation function is as follows:
\begin{equation}
\begin{split}
\Pi^{(\boldsymbol s)}(P,q)
&= i\!\int d^4x\, e^{i(P+q)\cdot x} \\
&\times
\langle0|
\mathcal T\{J_{B_d}(x)\mathcal O^{(\boldsymbol s)}(0)\}
|\Lambda(P)\rangle,
\end{split}
\label{eq:Correlator_s_def}
\end{equation}
where $\mathcal T$ is the time-ordering operator, $P$ represents the $\Lambda$'s momentum, and $q$ denotes the transferred momentum. Here the $B_d$-meson interpolating current is defined as $J_{B_d}(x)=\bar b(x)i\gamma_5 d(x)$. An analogous correlation function can be defined for the $(\boldsymbol b)$-model by replacing $\mathcal O^{(\boldsymbol s)}$ with $\mathcal O^{(\boldsymbol b)}$. The leading-order diagram corresponding to both correlation functions can be seen in Fig~\ref{fig:leadingDiagram}. In this diagram, $\Lambda$ enters through its DAs depicted by the shaded oval, while the $B$-meson enters through its interpolating current.

Within the framework of LCSR, the correlation function is evaluated from two perspectives: the physical (hadronic) side, where it is expressed in terms of hadronic parameters, and the QCD side, where it is expanded in terms of the $\Lambda$-baryon DAs ordered by increasing twist. In this region, the particle momenta are highly off-shell $(P+q)^2 \ll m^2_{b}$ and $q^2 \ll m^2_{b}$ and the integration in Eq.~\eqref{eq:Correlator_s_def} can be evaluated near the light-cone $x^2\sim 0$. In this approach, the perturbative amplitude is convoluted with the nonperturbative DAs of the $\Lambda$ baryon.

These two perspectives are equated by isolating the ground state and suppressing the contribution of the
higher states and continuum. To this end the Borel transformation and continuum subtraction will be applied to both sides of the correlation function, and the quark–hadron duality assumption will be imposed. The derivation of these two representations is discussed in the following subsections.
%%%%%%%%%%%%%%%%%%%%%%%%%%%%%%%%%%%%%%%

\subsection{PHYSICAL SIDE}
\label{subsec:physical_side}
To determine the physical side, a complete set of intermediate states with the same quantum numbers as $B_d$ meson is inserted into the correlation function. Using the dispersion relation analyticity, the correlator can then be written in terms of a dispersion relation. The integration of four-integrals leads to
\begin{equation}
\begin{aligned}
\Pi^{(\boldsymbol s)}(P,q)&= \frac1{m_{B_d}^2 - (P+q)^2}
\\&\times\langle 0 | J_{B_d}(0) | B_d(P+q) \rangle
\langle B_d(P+q) | \mathcal O^{(\boldsymbol s)} | \Lambda(P) \rangle
\\&
+ \int_{s_h}^\infty ds'
\frac{\rho^{(\boldsymbol s)}(s',P,q)}{s' - (P+q)^2},
\end{aligned}
\label{eq:Pi_dispersion}
\end{equation}
where $s_h$ denotes the lowest threshold of the excited states and continuum contributions, and $\rho^{(\boldsymbol s)}(s,P,q)$ represents the corresponding spectral density containing these contributions. 

The vacuum to $B_d$-meson matrix element of the interpolating current, $\langle0|J_{B_d}(0)|B_d(P+q)\rangle$, can be written in terms of the $B_d$-meson decay constant $f_{B_d}$:
 \begin{equation}
  \langle 0 | J_{B_d}(0) | B_d(P+q) \rangle =\frac{m_{B_d}^2 f_{B_d}}{m_b+m_d}.
  \label{eq:decay-constant}
 \end{equation} 
For the rest of this work, we neglect $m_d$ in comparison to the $b$-quark mass. The correlator in Eq.~\eqref{eq:Pi_dispersion} can be decomposed into four invariant amplitudes, corresponding to independent Dirac structures:
\begin{equation}
    \begin{aligned}
        \Pi^{(\boldsymbol s)}(P,q) &=  \Pi_R^{(\boldsymbol s)}((P+q)^2, q^2) u_{\Lambda,R}(P)
        \\ &+ \Pi_L^{(\boldsymbol s)}((P+q)^2, q^2) u_{\Lambda,L}(P)                              \\ & + \tilde{\Pi}_R^{(\boldsymbol s)}((P+q)^2, q^2)\, \slashed{q}\, u_{\Lambda,R}(P)
        \\ &+ \tilde{\Pi}_L^{(\boldsymbol s)}((P+q)^2, q^2)\, \slashed{q}\, u_{\Lambda,L}(P),
    \end{aligned}
    \label{eq:Pi_decomposition}
\end{equation}
where $u_{\Lambda,R(L)}$ is the right (left)-handed $\Lambda$ spinor. Contributions involving $\slashed P u_{\Lambda,R(L)}$ are accounted for in the first two terms due to the Dirac equation. Using this decomposition, each invariant amplitude can be expressed in terms of a dispersion relation which involves the corresponding form factor. For example, for $\Pi^{(\boldsymbol s)}_R$ one obtains:
\begin{equation}
    \begin{aligned}
        \Pi_R^{(\boldsymbol s)}((P+q)^2, q^2) & =  \frac{m_{B_d}^2 f_{B_d}}{m_b} \frac{F_{B_d\to\Lambda_R}^{(\boldsymbol s)}(q^2)}{m_{B_d}^2 - (P+q)^2} \\ & + \int_{s_h}^\infty ds' \frac{\rho_R^{(\boldsymbol s)}(s',q^2)}{s' - (P+q)^2},
    \end{aligned}
    \label{eq:dispersion_R_final}
\end{equation}
 and for $\tilde\Pi^{(\boldsymbol s)}_L(q^2)$, the corresponding relation is
 \begin{equation}
    \begin{aligned}
       \tilde \Pi_L^{(\boldsymbol s)}((P+q)^2, q^2) & =  \frac{m_{B_d}^2 f_{B_d}}{m_b m_\Lambda} \frac{\tilde F_{B_d\to\Lambda_L}^{(\boldsymbol s)}(q^2)}{m_{B_d}^2 - (P+q)^2} \\ & + \int_{s_h}^\infty ds' \frac{\tilde\rho_L^{(\boldsymbol s)}(s',q^2)}{s' - (P+q)^2}.
    \end{aligned}
    \label{eq:dispersion_L_final}
\end{equation}
Similar relations hold for $\Pi_L^{(\boldsymbol s)}((P+q)^2, q^2)$ and $\tilde\Pi_R^{(\boldsymbol s)}((P+q)^2, q^2)$. The corresponding form factors for model $(\boldsymbol b)$ can be derived in an entirely analogous manner. Throughout this work, we employ the shorthand notation $F_{R(L)}$  and $\tilde F_{R(L)}$ to denote the form factors introduced above appearing in both models. 
%%%%%%%%%%%%%%%%%%%%%%%%%%%%%%%%%%%%%%%%%%%%%%%%%%%%%%%%%%%

\subsection{QCD SIDE}
\label{subsec:QCD_side}
In this subsection, the correlator of each model is calculated in deep Euclidean region via the OPE near the light cone. Equations.~\eqref{eq:O_s} and~\eqref{eq:O_b} involving the $B_d$-meson interpolating current and the local operators governing the interaction are expressed in terms of quark fields. The field contractions are performed using the Wick's theorem, yielding an expression in terms of the heavy‑quark propagator and light‑quark fields sandwiched between the vacuum and the $\Lambda$-baryon state. The correlation function for the $(\boldsymbol s)$-model then takes the form
\begin{equation}
    \begin{aligned}
        \Pi_{\eta}^{(\boldsymbol s)}(P,q)& =
        -i\int  d^4x \, e^{i(P+q)\cdot x}
        (P_R)_{\eta\gamma}
        \big[ C P_R S_b(-x)\gamma_5 \big]_{\alpha\beta}
        \\ & \times \epsilon_{ijk}
        \langle 0 |u_\alpha^i(0) d_\beta^j(x) s_\gamma^k(0) | \Lambda(P) \rangle,
    \end{aligned}
    \label{eq:QCD:wick:s}
\end{equation}
and for the $(\boldsymbol b)$-model, the corresponding correlator becomes
\begin{equation}
    \begin{aligned}
        \Pi_{\eta}^{(\boldsymbol b)}(P,q)& =
        i\int  d^4x \, e^{i(P+q)\cdot x}
        \big[ P_R S_b(-x) \gamma_5 \big]_{\eta\beta}
        \big[ C P_R \big]_{\gamma\alpha} \\ &
        \times\epsilon_{ijk}
        \langle 0 | u_\alpha^i(0) d_\beta^j(x) s_\gamma^k(0) | \Lambda(P) \rangle.
    \end{aligned}
    \label{eq:QCD:wick:b}
\end{equation}
Here $\alpha, \beta, \gamma, \eta$ denote matrix indices, $i, j, k$ are color indices, and $C$ indicates the charge-conjugation operator. The heavy‑quark propagator is defined by $S_b$, and contributions proportional to quark and gluon condensates are neglected. The matrix index $\eta$ is carried by the $\Lambda$ spinor in the physical side in Eq.~\eqref{eq:Pi_decomposition}, implicitly.

The short distance effects are calculated within perturbative QCD, whereas the long distance effects, i.e., the nonperturbative information, is encoded in the matrix element appearing in Eqs.~\eqref{eq:QCD:wick:s} and~\eqref{eq:QCD:wick:b}, which are decomposed into a set of DAs of increasing twist, defined in terms of the light-cone variable $x$. Following the standard parameterization for octet baryons~\cite{Liu:2008yg,Aliev:2010uy,Liu:2009mb,Liu:2009sn}, these matrix elements are expanded as:
\begin{equation}
    \epsilon_{ijk} \langle 0 | u_\alpha^i(a_1 x) d_\beta^j(a_2 x) s_\gamma^k(a_3 x) | \Lambda(P) \rangle = \sum_i \mathcal{F}_i(x) \Gamma_i,
    \label{eq:DA_decomposition}
\end{equation}
where $\mathcal{F}_i$ are the invariant functions involving the DAs (such as $\mathcal{V}_1, \mathcal{A}_1, \mathcal{T}_1$, etc., up to twist-6) and $\Gamma_i$ are the corresponding Dirac structures; see Appendix~\ref{sec:appA}. For the matrix elements in Eqs.~\eqref{eq:QCD:wick:s} and \eqref{eq:QCD:wick:b}, $a_1=a_3=0$ and $a_2=1$. After substituting Eqs.~(\ref{eq:DAs}--\ref{eq:QCD:DA:generic}) into the correlators for each model, powers of the scalar product $P\cdot x$ appear in the denominator. A partial integration with respect to the dimensionless variable $x_i$ (see Eq.~\eqref{eq:QCD:DA:generic}) is introduced for each power of $P\cdot x$ and the surface terms vanish. The four-dimensional integrals over the momentum of the $b$-quark propagator and $x$ are then performed. We find that all terms containing $T_i$ vanish in model $(\boldsymbol s)$. 

At this stage, in order to suppress the contributions of higher-resonance and continuum states, a Borel transformation with respect to $(P+q)^2$, followed by continuum subtraction, is applied to the physical and QCD sides of the correlation functions in Eqs.~(\ref{eq:dispersion_R_final},~\ref{eq:dispersion_L_final}) and~(\ref{eq:QCD:wick:s},~\ref{eq:QCD:wick:b}), respectively. After applying the Borel replacement rules given in Eqs.~(\ref{eq:Borel}--\ref{eq:x0_definition}) to both sides of the correlation function and invoking the quark-hadron duality assumption component by component on physical side and QCD side, the final light‑cone sum rule for each form factor can be extracted. The right-handed and left-handed form factors in the $(\boldsymbol s)$-model are obtained from the following relations:
\begin{subequations}
\begin{equation}
\begin{aligned}
    \mathcal{B}_{M^2} \Pi_R^{( s)}(Q^2,M^2,s_0) &= \frac{m_{B_d}^2}{m_b} f_{B_d} e^{-m_{B_d}^2/M^2} \\& \times F_R^{(\boldsymbol s)}(Q^2, M^2, s_0),
 \end{aligned}   
\end{equation}
\begin{equation}
\begin{aligned}
    \mathcal{B}_{M^2} \tilde{\Pi}_L^{(\boldsymbol s)}(Q^2,M^2,s_0) &=-\frac{m_{B_d}^2}{m_b m_\Lambda} f_{B_d} e^{-m_{B_d}^2/M^2} \\& \times \tilde F_L^{(\boldsymbol s)}(Q^2, M^2, s_0),
    \end{aligned}
\end{equation}
\label{eq:FF_Borel}
\end{subequations}
where $\mathcal{B}_{M^2}$ denotes the Borel transformation and continuum subtraction. Similar relations hold for the correlation functions $\Pi_R^{(\boldsymbol b)}$ and $\tilde{\Pi}_L^{(\boldsymbol b)}$. The explicit expressions for the left-hand sides of the above equations for both models are provided in Appendix~\ref{sec:appC}.
%%%%%%%%%%%%%%%%%%%%%%%%%%%%%%%%%%%%%%%%%%%%%%%%%%%%%%%%%%%
\section{Numerical analysis}
\label{sec:Numerical_analysis}
In this section, we analyse the derived form factors and the resulting branching fraction numerically. The light-cone sum rules provide reliable results only for the region $(P+q)^2,q^2\ll m_{b}^2$. However, the physical decay $B_d \to \Lambda \psi$ spans a kinematic range up to $(m_{B_d} - m_\Lambda)^2$, extending into the timelike region where the OPE becomes unreliable. To extrapolate our results to the full physical domain and obtain analytic expressions for the form factors, we employ the method of $z$-expansion \cite{Boyd:1995cf,Bourrely:2008za,Khodjamirian:2017fxg,Ahmadi:2025oal}.

The numerical analysis requires several input parameters, including hadronic and QCD quantities, variables entering the $\Lambda$-baryon DAs, auxiliary parameters associated with the LCSR framework, and the couplings appearing in each model. Some of these input parameters are summarized in Table~\ref{tab:inputs}; see also Refs.~\cite{ParticleDataGroup:2024cfk, Liu:2008yg, Lenz:2024rwi}. The upper panel lists the $B_d$-meson mass and decay constant, the $b$-quark mass in the $\overline{\mathrm{MS}}$ scheme, and the $\Lambda$-baryon mass. The middle panel shows the parameters relevant to the $\Lambda$ distribution amplitudes, while the lower panel contains the estimated couplings for the models considered. We have adopted some parameters for the mass correction terms $\mathcal V_1^M$, $\mathcal A_1^M$ and $\mathcal T_1^M$ from the Nucleon analysis, since no dedicated $\Lambda$ determination of them exists in the literature. These dimensionless parameters are $V_1^u=0.23\pm0.03$, $A_1^M=0.38\pm0.03$, $f_1^u=0.07\pm0.05$, $f_1^d=0.40\pm0.05$ and $f_2^d=0.22\pm0.05$. Because in the matrix elements in Eqs.~\eqref{eq:QCD:wick:s} and~\eqref{eq:QCD:wick:b} the $d$-quark field carries the nonzero light-cone separation $x$, all terms containing $\mathcal A_1^M$ vanish\cite{Braun:2006hz}. The resulting $SU(3)$-symmetry approximation has a negligible impact on our numerical results. For example, the effect of these parameters in $F_R^{(\boldsymbol s)}(0)$ is about $1\%$.
\begin{table}[tb]
    \begin{ruledtabular}
        \begin{tabular}{|cc|}
            Parameter                     & Interval                         \\
            \colrule
            $m_\Lambda$                   & $1.115683 \pm 0.000006~\text{GeV}$   \\
            $m_{B_d}$                         & $5.27972 \pm 0.00008~\text{GeV}$    \\
            $m_b(\overline{\mathrm{MS}})$ & $4.183 \pm 0.007~\text{GeV}$         \\
            $f_{B_d}$                         & $0.189 \pm 0.0014~\text{GeV}$       \\
            \colrule
            $f_\Lambda$  & $(6.0\pm0.3)\times10^{-3}~\text{GeV}^2$         \\
            $\lambda_1$  & $(1.0\pm0.3)\times10^{-2}~\text{GeV}^2$             \\
            $|\lambda_2|$ & $(0.83\pm0.05)\times 10^{-2}~\text{GeV}^2$\\
            $|\lambda_3|$ &$(0.83\pm0.05)\times 10^{-2}~\text{GeV}^2$\\
            \colrule
            $|G^{(\boldsymbol s)}|^2$                     & $1 \times 10^{-13}~\text{GeV}^{-4} $ \\
            $|G^{(\boldsymbol b)}|^2$                     & $3.7 \times 10^{-14}~\text{GeV}^{-4}$\\
        \end{tabular}
        \caption{The numerical values of the main input parameters.}
        \label{tab:inputs}
    \end{ruledtabular}
\end{table}
\subsection{Form Factors}
\label{subsec:FFs}
Furthermore, to obtain reliable form factor predictions, appropriate working regions for the auxiliary parameters must be specified. These include the Borel parameter $M^2$, and the continuum threshold $s_0$. Constraints arising from the QCD side calculation restrict the allowed range of  $s_0$. In particular, the requirement that the expression under the square root in Eq.~\eqref{eq:x0_definition} be positive yields two possible intervals for $Q^2$. From the analyses, we get $x_0$ to be within the interval $(0,1)$, which leads to the following ranges for $s_0$:
\[
    m_b^2  < s_0 \le m_b^2 + m_\Lambda^2, \qquad s_0 > m_b^2 + m_\Lambda^2 .
    \label{eq:x0_conditions}
\]
The first interval is extremely low to be the threshold of the $B_d$-meson first excited states and continuum contributing to the correlation function, and therefore the second interval is adopted in the numerical analysis. 

\begin{figure*}[tb]
  \centering
  \subfloat[]{\includegraphics[width=0.48\textwidth]{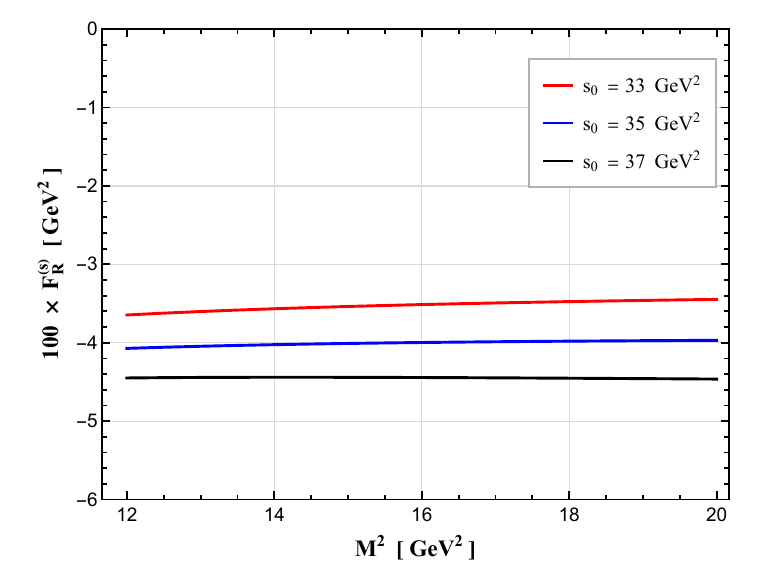}}
  \subfloat[]{\includegraphics[width=0.48\textwidth]{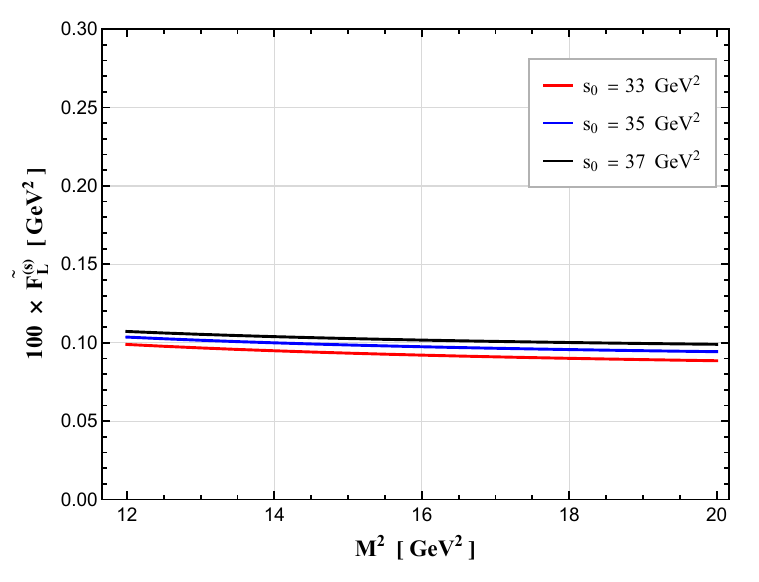}}\\
  \subfloat[]{\includegraphics[width=0.48\textwidth]{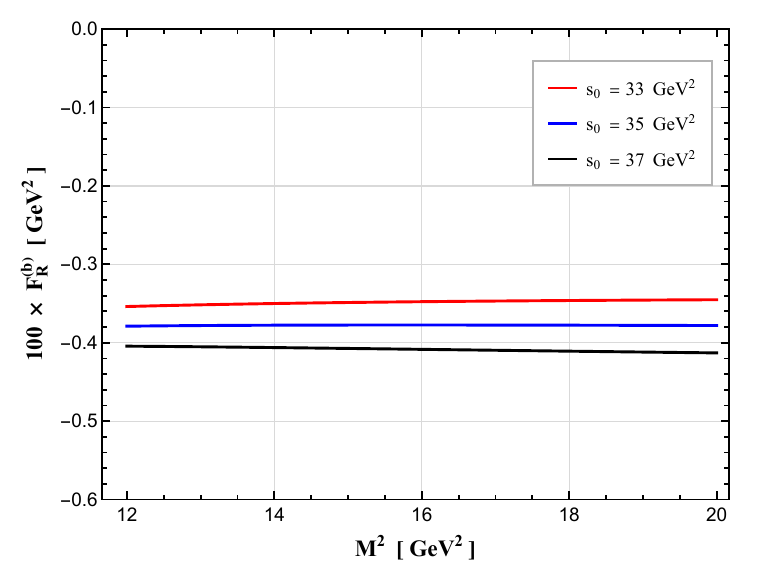}}
  \subfloat[]{\includegraphics[width=0.48\textwidth]{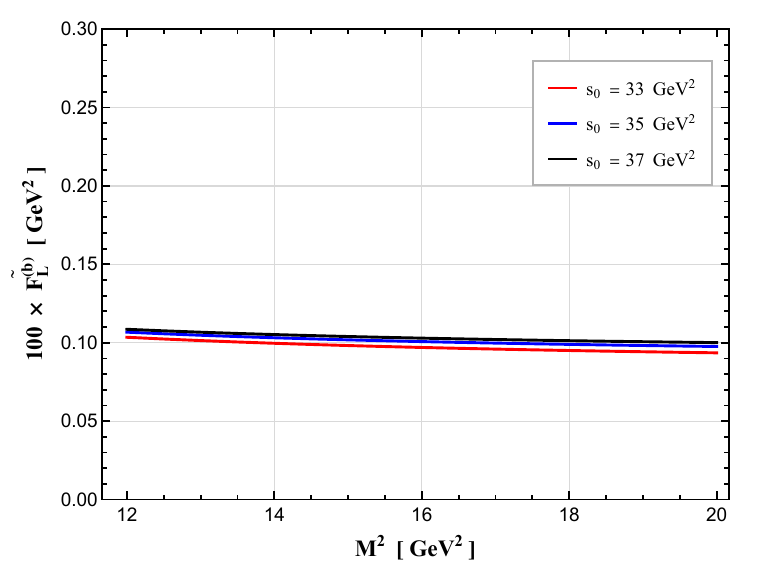}}
  \caption{Dependence of the form factors on the Borel parameter $M^2$ at $Q^2=1\ \mathrm{GeV}^2$ for different values of the continuum threshold $s_0$ within the chosen working region.
  (a) The right-handed form factor $F_R^{(\boldsymbol{s})}$;
  (b) The left-handed form factor $\tilde F_L^{(\boldsymbol{s})}$;
  (c) The right-handed form factor $F_R^{(\boldsymbol{b})}$;
  (d) The left-handed form factor $\tilde F_L^{(\boldsymbol{b})}$.}
  \label{fig:Msq}
\end{figure*}

\begin{figure*}[t]
  \centering
  \subfloat[]{\includegraphics[width=0.48\textwidth]{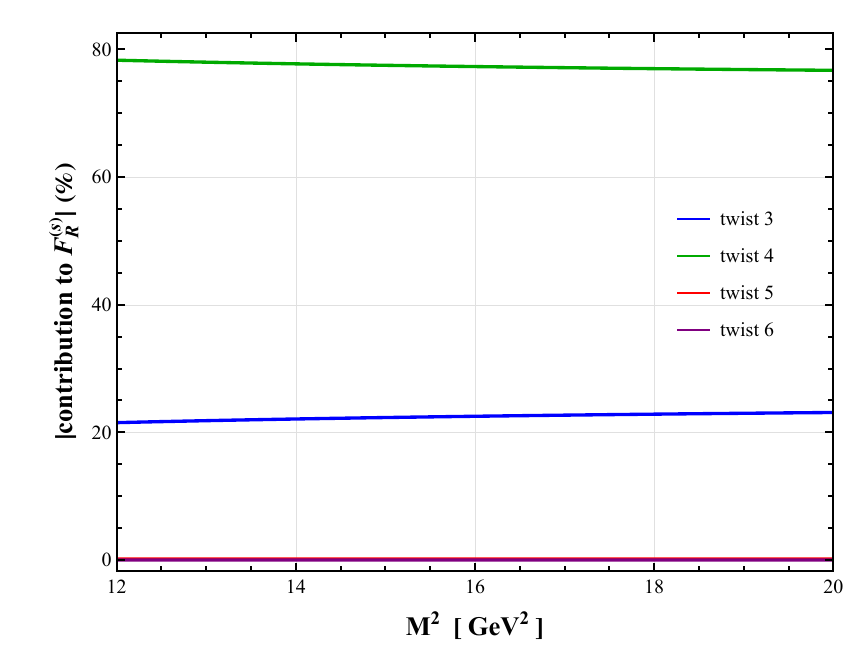}}
  \subfloat[]{\includegraphics[width=0.48\textwidth]{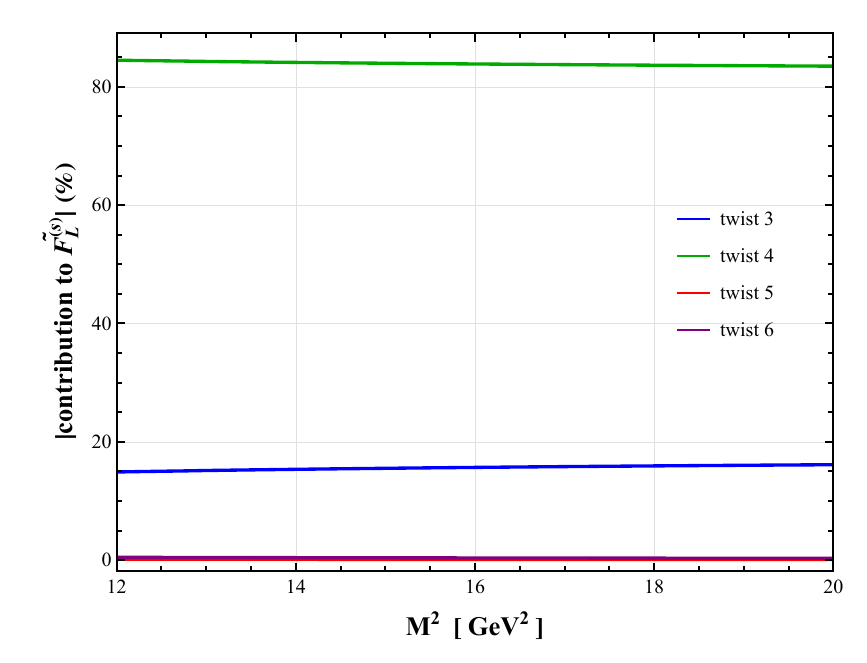}}\\
  \subfloat[]{\includegraphics[width=0.48\textwidth]{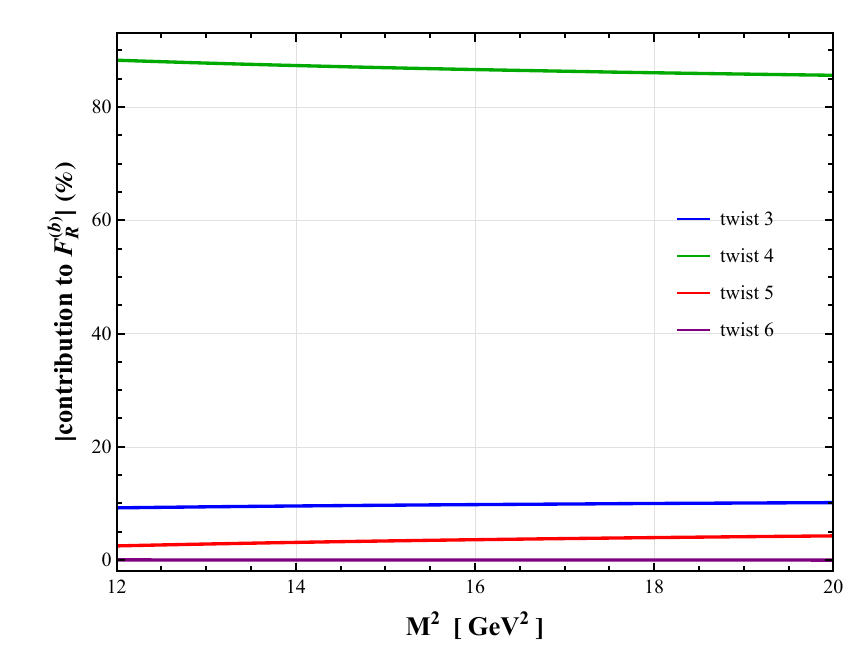}}
  \subfloat[]{\includegraphics[width=0.48\textwidth]{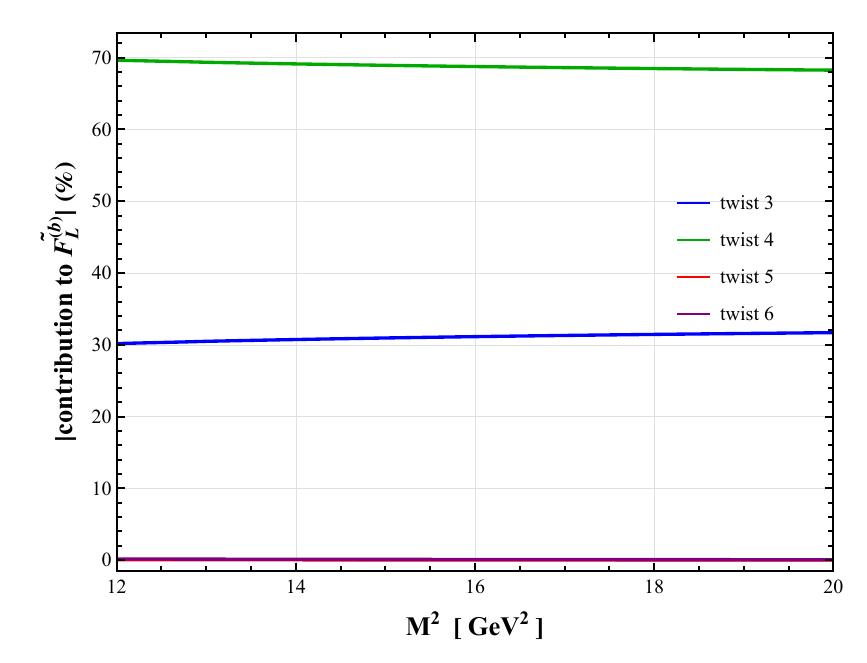}}
  \caption{Percentage contributions of different twists to the form factors as a function of the Borel parameter $M^2$ at $s_0=35\ \mathrm{GeV}^2$ and $Q^2=1\ \mathrm{GeV}^2$. Each curve gives the magnitude of a given twist's contribution as a percentage of the summed absolute contributions of all twists.
  (a) $F_R^{(\boldsymbol s)}$;
  (b) $\tilde F_L^{(\boldsymbol s)}$;
  (c) $F_R^{(\boldsymbol b)}$;
  (d) $\tilde F_L^{(\boldsymbol b)}$.}
  \label{fig:twistCont}
\end{figure*}

The standard QCD sum rule approach requires that the pole contribution dominate over the contributions of higher resonances and the continuum, which imposes an upper bound on the working range of the Borel parameter. At the same time, the lower bound of the interval is determined by requiring the dominance of the contributions from lower‑twist $\Lambda$ DAs over those of higher twists. Accordingly, the Borel window is chosen such that the resulting form factors in each model exhibit only a weak dependence on $M^2$. The pole contribution $PC$ and the convergence parameter $R$ for the right-handed correlation function of the ($\boldsymbol s$) model are defined as 
\begin{equation}
\begin{split}
&PC^{(\boldsymbol s)}_R=\frac{\mathcal B_{M^2}\Pi_R^{(\boldsymbol s)}(Q^2,M^2,s_0)}{\mathcal B_{M^2}\Pi_R^{(\boldsymbol s)}(Q^2,M^2,\infty)}\\& R^{(\boldsymbol s)}_R=\frac{\mathcal B_{M^2}\Pi_R^{(\boldsymbol s),(\text{highest twist})}(Q^2,M^2,s_0)}{\mathcal B_{M^2}\Pi_R^{(\boldsymbol s)}(Q^2,M^2,s_0)}\label{eq:PCR}
\end{split}
\end{equation}
Following the standard QCD sum rule criteria, we require that $PC>50\%$ and $R<5\%$. These conditions lead to the working regions for these two parameters that is also in agreement with the $B_d$ meson two-point correlation function; see also Ref.~\cite{Shifman:1978bx}. For example, at $M^2=16\ \mathrm{GeV}^2$, $s_0=35\ \mathrm{GeV}^2$ and $Q^2=1\ \mathrm{GeV}^2$, the pole contribution for the right-handed correlation function of model ($\boldsymbol s$) is $51.5\%$ and the twist-$6$ contribution for the same function amounts to $0.006\%$ of the total. The chosen intervals are as follows:
\[
    s_0 \in [33,37]\ \mathrm{GeV}^2,
    \qquad
    M^2 \in [12,20]\ \mathrm{GeV}^2.
\]
As seen in Fig.~\ref{fig:Msq}, at $Q^2 = 1\ \mathrm{GeV}^2$ and within the chosen Borel window, the two form factors in each model exhibit negligible dependence on $M^2$. Also, Fig.~\ref{fig:twistCont} shows the relative twist contributions to $F_R^{(\boldsymbol{s})}, \tilde F_L^{(\boldsymbol s)}, F_R^{(\boldsymbol b)} \, \text{and} \, \tilde F_L^{(\boldsymbol b)}$ at $Q^2=1~\mathrm{GeV}^2$ and $s_0=35~\mathrm{GeV}^2$ as a function of the Borel parameter within the chosen interval. It can be seen from the figure that the dependence of each twist contribution on Borel parameter $M^2$ is negligible in the chosen working region for all the form factors. While the twist-5 and twist-6 contributions remain suppressed compared with the lower-twist contributions, it is evident that the twist-4 contribution is numerically larger than the twist-3, twist-5 and twist-6 contributions for all the four form factors in both models. 

This behavior is contrary to the $B^+ \to p$ study of higher twists in Ref.~\cite{Boushmelev:2023huu}, where their twist-3 is larger than twist-4 in their $(\boldsymbol d)$-model (closest to our $(\boldsymbol s)$-model), since the tensor distribution amplitudes $(T_i)$s vanish identically in both form factors, while the twist hierarchy is interchanged only in their $(\boldsymbol b)$-model. In the mentioned study, they link this behavior to sizable $T_1, T_2, T_3 \, \text{and}\, T_7$ contributions which are absent in $(\boldsymbol d)$-model. 

Considering that $(T_i)$s vanish identically in our $(\boldsymbol s)$-model, this particular reason cannot explain the twist-4 dominance we observe here. Instead, we find that using the $\Lambda$ DAs, the twist-3 contribution to the $(\boldsymbol s)$-model sum rule arises only from $A_1$ (since $V_1 \,\text{and} \, T_1$ are zero for $\Lambda$-baryon), whereas twist-4 contribution comes from $S_1, P_1, V_3, A_2 \, \text{and}\, A_3$ (with $V_2, T_2, T_3 \,\text{and}\, T_7$ vanishing). This can be the origin of the observed dominance of twist-4; however, further investigation would require for higher accuracy distribution amplitudes of $\Lambda$-baryon including higher twists, especially from lattice QCD.


The extraction of reliable form factors using the $z$-expansion is now in order. The BCL version of the $z$-expansion \cite{Bourrely:2008za},  involves mapping the complex $q^2$-plane onto the unit disk in the $z(q^2)$-plane. The mapping is defined as:
\begin{equation} 
        z(q^2)  = \frac{\sqrt{t_+ - q^2} - \sqrt{t_+ - t_0}}{\sqrt{t_+ - q^2} + \sqrt{t_+ - t_0}}, ,
\end{equation}
where
\begin{equation}
    \begin{aligned}
          t_{\pm}&=(m_{B_d} \pm m_\Lambda)^2,
           \\  t_{0}&=(m_{B_d} + m_\Lambda)(\sqrt{m_{B_d}} - \sqrt{m_\Lambda})^2.
    \end{aligned}
\end{equation}
Here, $t_{-}$ represents the physical upper bound on $q^2$ such that the two-body decay remains kinematically allowed, $t_{+}$ corresponds to the threshold for multiparticle states and higher resonances, and $t_0$ is chosen such that the point $q^2=t_0$ is mapped onto the origin of the complex plane, i.e., $z(t_0)=0$. The form factors are then parameterized as:
\begin{equation}
    \begin{aligned}
        F  (q^2) &=  \frac{F(0)}{1 - q^2/m_{\Lambda_b}^2}
        \\&\times\biggl[ 1 + b \biggl(z(q^2) - z(0)
            + \frac{1}{2}\bigl[z(q^2)^2 - z(0)^2\bigr]\biggr)\biggr],
    \end{aligned}
    \label{eq:zexp_final}
\end{equation}
where $F(0)$ and $b$ are fit parameters. The choice of $t_{+}$ introduces an isolated pole at $q^2=m_{\Lambda_b}^2$ inside the unit disk of the $z$-plane, corresponding to the lowest-lying resonance with the appropriate quantum numbers in this channel. The parameter $F(0)$ denotes the value of the form factor at $q^2=0$, while $b$ characterizes the slope of the form factor.

To extract the fit parameters, Eq.~\eqref{eq:zexp_final} is fitted to the LCSR results obtained in the region
\[
-10~\mathrm{GeV}^2 \le q^2 \le 0~\mathrm{GeV}^2 .
\]
This interval ensures a stable extrapolation of the form factors to the high-$q^2$ region. The resulting fit parameters for the $(\boldsymbol{s})$ and $(\boldsymbol{b})$ models are summarized in Table~\ref{tab:fit_params}. Their uncertainties are obtained by varying the auxiliary parameters $M^2$ and $s_0$
over their working windows, together with the input parameters within the errors quoted
in Table~\ref{tab:inputs}. Each parameter is varied independently, with the remaining
ones held at their central values, and the form factors are refitted for every
variation. The two auxiliary parameters $M^2$ and
$s_0$ are treated as uncorrelated.

The LCSR form factors evaluated directly at $q^2=0$, prior to the $z$-expansion extrapolation, are presented below for direct comparison with the fitted values $F(0)$ in Table~\ref{tab:fit_params}:
\begin{itemize}
\item model ($\boldsymbol s$):
\[
\begin{split}
&F_R^{(\boldsymbol s)}(0)\times10^2=-4.311\pm0.427,\\& \tilde F_L^{(\boldsymbol s)}(0)\times10^2=0.107\pm0.006.
\end{split}
\]
\item model ($\boldsymbol b$):
\[
\begin{split}
&F_R^{(\boldsymbol b)}(0)\times10^2=-0.400\pm0.030,\\& \tilde F_L^{(\boldsymbol b)}(0)\times10^2=0.109\pm0.005.
\end{split}
 \]
\end{itemize}
These values are in good agreement with the fitted $F(0)$ parameters of Table~\ref{tab:fit_params} confirming that the $z$-expansion parametrization faithfully reproduces the direct LCSR results at $q^2=0$.
\begin{table}[tb]
    \begin{ruledtabular}
        \begin{tabular}{|ccc|}
            \toprule
                         & $F(0)\times 10^{2}$                           & $b$                    \\
            \colrule
            $F_R^{(\boldsymbol s)}(q^2)$        & $-4.410 \pm 0.610$           & $- 7.930 \pm 0.300$ \\ 
            $\tilde F_L^{(\boldsymbol s)}(q^2)$ & $0.105 \pm 0.012$ & $-9.190 \pm 0.490$ \\
            \colrule
            $F_R^{(\boldsymbol b)}(q^2)$        & $-0.405 \pm 0.041$           & $-3.820 \pm 1.000$ \\ 
            $\tilde F_L^{(\boldsymbol b)}(q^2)$ & $0.108 \pm 0.009$ & $-5.390 \pm 0.83$
            \\
        \end{tabular}
        \caption{Fit parameters of the $z$-expansion for the $(\boldsymbol s)$ and $(\boldsymbol b)$ models.}
        \label{tab:fit_params}
    \end{ruledtabular}
\end{table}

\begin{figure*}[tb]
  \centering
  \subfloat[]{\includegraphics[width=0.48\textwidth]{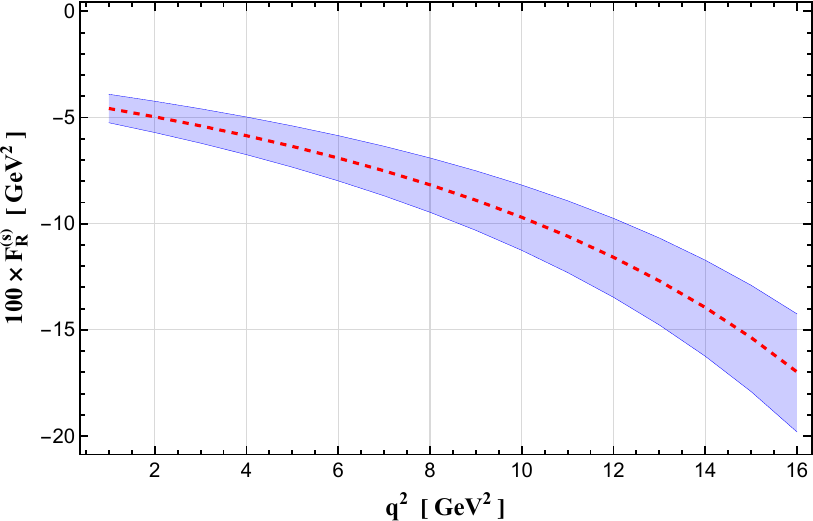}}
  \subfloat[]{\includegraphics[width=0.48\textwidth]{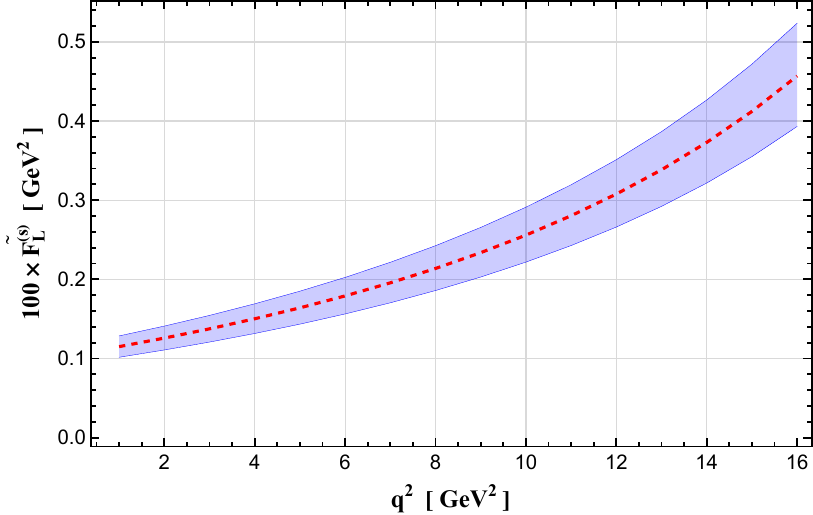}}\\
  \subfloat[]{\includegraphics[width=0.48\textwidth]{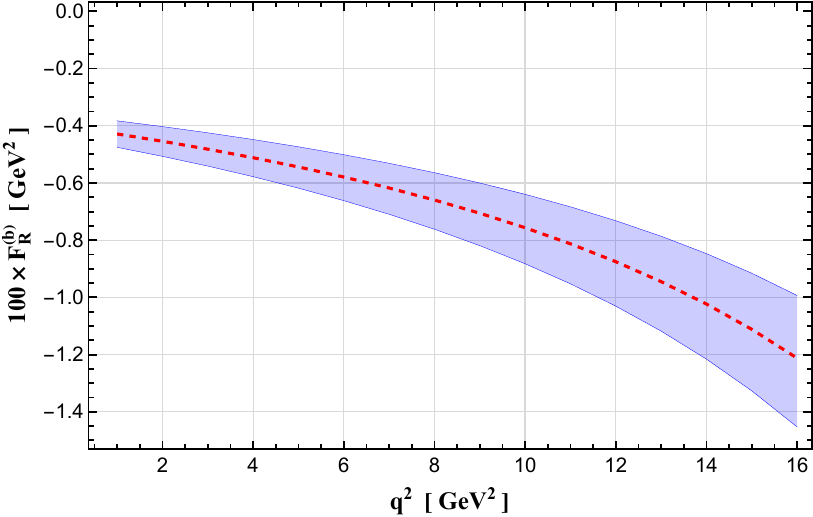}}
  \subfloat[]{\includegraphics[width=0.48\textwidth]{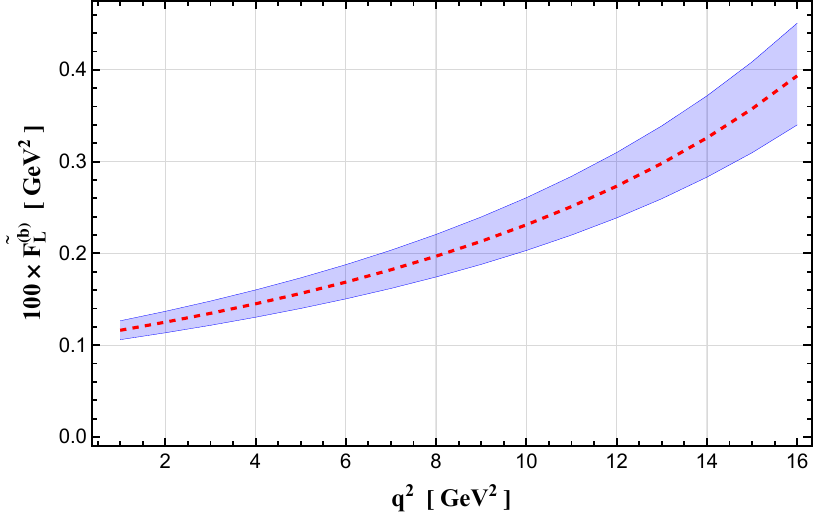}}
  \caption{Fitted form factors for the decay mode $B_d\to\Lambda\psi$ as functions of $q^2$.
  (a) $F_R^{(\boldsymbol{s})}$;
  (b) $\tilde F_L^{(\boldsymbol{s})}$;
  (c) $F_R^{(\boldsymbol{b})}$;
  (d) $\tilde F_L^{(\boldsymbol{b})}$. The shaded bands represent the total theoretical uncertainty, obtained by varying the Borel parameter
  $M^2$ and the continuum threshold $s_0$ within their working windows together with the
  uncertainties of the input parameters listed in Table~\ref{tab:inputs}.}
  \label{fig:FF}
\end{figure*}

The numerical results for the form factors obtained from the $z$-expansion are presented in Figure~\ref{fig:FF} for the $(\boldsymbol{s})$ and $(\boldsymbol{b})$ models. The uncertainty bands reflect the variation of the auxiliary parameters $s_0$ and $M^2$ within their working regions, together with the uncertainties of the input parameters given in Table~\ref{tab:inputs}. It should be noted that the next-to-leading-order (NLO) QCD corrections to the sum rules can further impact the uncertainty bands, as Ref.~\cite{Wang:2015ndk} shows that these corrections would reduce the tree-level predictions for heavy to light baryon transition from factors up to $\mathcal{O}(50 \%)$.   
%%%%%%%%%%%%%%%%%%%%%%%%%%%%%%%%%%%%%%%%%%%%%%%%%%%%%%%%%%%%%%%%%%%%%%
\subsection{Branching Fractions}
\label{subsec:Br}
\begin{figure*}[tb]
  \centering
    \subfloat[model $(\boldsymbol{s})$]{\includegraphics[width=0.48\textwidth]{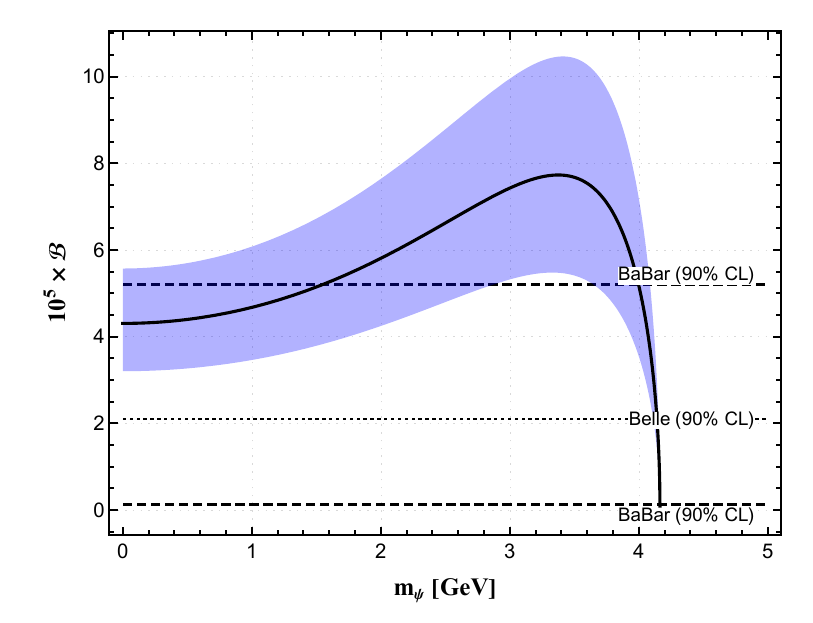}}
    \subfloat[model $(\boldsymbol{b})$]{\includegraphics[width=0.48\textwidth]{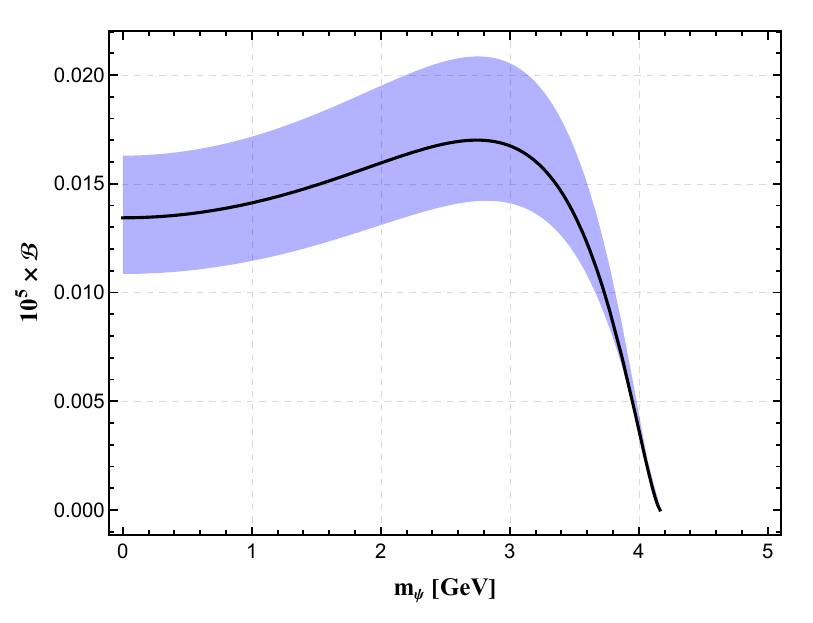}}
  \caption{
    The branching fractions for the process $B_d \to \Lambda\,\psi$ mediated by the operators 
    $\mathcal{O}^{(\boldsymbol{s})}$ and $\mathcal{O}^{(\boldsymbol{b})}$ as functions of the dark-sector
    antibaryon mass $m_\psi$. The uncertainty bands arise from the uncertainties in the fit parameters 
    $F(0)$ and $b$ as well as the input hadron masses. The horizontal lines indicate the current 
    experimental upper limits from BaBar and Belle.
  }
  \label{fig:Br}
\end{figure*}

The expression for the two-body decay width of the process $B_d \to \Lambda\psi$ is given in Eq.~\eqref{eq:two_body_decay_final}. In each model, only two form factors contribute. Equation~\eqref{eq:two_body_decay_final} for the $(\boldsymbol{s})$ model can therefore be written in terms of these form factors as
\begin{equation}
\begin{aligned}
\Gamma^{(\boldsymbol s)}(B_d\to\Lambda\psi)&= \frac{|G^{(\boldsymbol s)}|^2}{16\pi m_{B_d}^3} \Bigg\{\Bigg[\frac{m_\psi^2}{m_\Lambda^2}\Big(\tilde F_L^{(\boldsymbol s)}(m_\psi^2)\Big)^2
\\&+\Big(F_R^{(\boldsymbol s)}(m_\psi^2)\Big)^2\Bigg]\Big(m_{B_d}^2-m_\Lambda^2-m_\psi^2\Big)
\\&+4m_\psi^2 F_R^{(\boldsymbol s)}(m_\psi^2)\tilde F_L^{(\boldsymbol s)}(m_\psi^2)\Bigg\}
\\&\times \lambda^{1/2}(m_{B_d}^2,m_\Lambda^2,m_\psi^2).
\label{eq:decayW2}
\end{aligned}
\end{equation}
The corresponding expression for the $(\boldsymbol{b})$ model is obtained by replacing $(\boldsymbol{s})$ with $(\boldsymbol{b})$. The branching fractions in these models are of particular interest, since experimental upper limits are available for comparison. By multiplying Eq.~\eqref{eq:decayW2} by the $B_d$-meson lifetime $\tau_{B_d}=1.517\pm0.004\,\mathrm{ps}$~\cite{ParticleDataGroup:2024cfk}, we obtain the branching fraction of each model as a function of the dark antibaryon mass $m_\psi$. The numerical analysis of the branching fractions is performed using the form factors obtained from the $z$-expansion extrapolation. The resulting branching fractions for the $(\boldsymbol{s})$ and $(\boldsymbol{b})$ models are displayed in Fig.~\ref{fig:Br}. 

The branching fractions exhibit an initial rise followed by a decline as the dark antibaryon mass increases, eventually vanishing at the kinematic endpoint value $m_\psi = 4.16\,\mathrm{GeV}$. This value corresponds to the difference between the $B_d$-meson and $\Lambda$ masses, as required by kinematics.

\begin{figure*}[tb]
  \centering
    \subfloat[model $(\boldsymbol{s})$]{\includegraphics[width=0.48\textwidth]{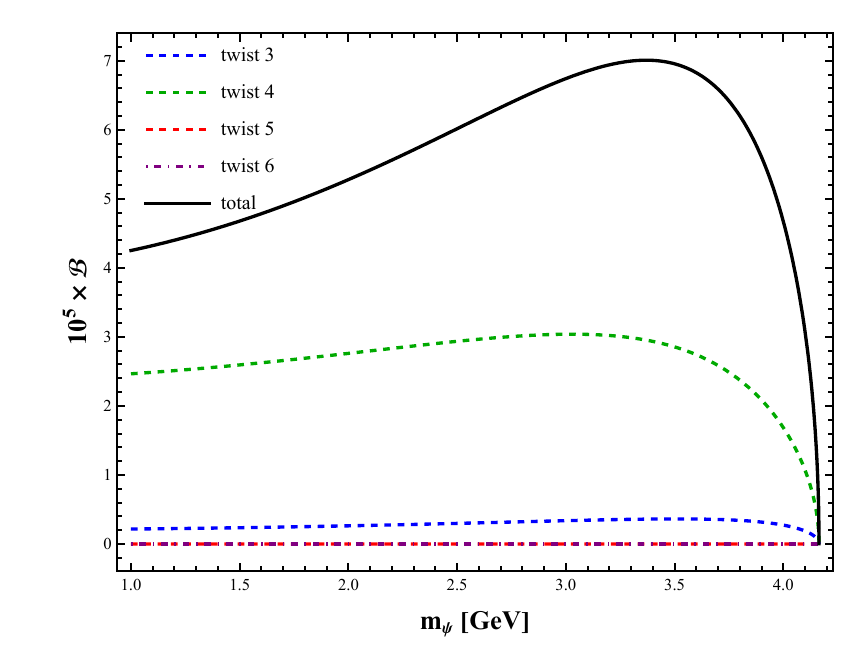}}
    \subfloat[model $(\boldsymbol{b})$]{\includegraphics[width=0.48\textwidth]{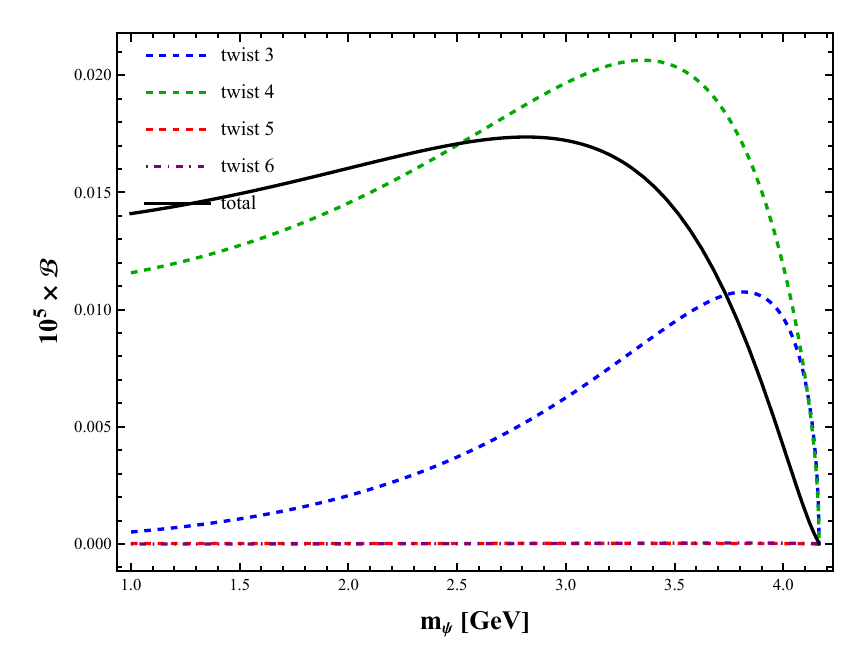}}
  \caption{Twist contributions to the branching fraction $\mathcal{B}(B_d \to \Lambda\,\psi)$ as a function of the dark antibaryon mass for the $(\boldsymbol s)$ model (a) and the $(\boldsymbol b)$ model (b). The black curve identifies the full result with DAs up to twist-6, while each colored curve denotes the branching fraction computed using only a single twist.}
  \label{fig:Br_twists}
\end{figure*}

It is instructive to show how each twist contributes to the branching fraction. Fig.~\ref{fig:Br_twists} presents the twist-by-twist decomposition of $\mathcal{B}(B_d\to\Lambda\psi)$ for the two models. As evident from the figure, the twist-4 and twist-3 $\Lambda$ DAs supply the main contributions to the branching fraction, whereas the twist-5 and twist-6 contributions are negligible over the whole mass range. It should be noted that since Eq.~\eqref{eq:decayW2} is quadratic in the form factors, single twist curves do not add up linearly to the total result.

The upper limits reported by the BaBar and Belle collaborations on the branching fraction of the decay $B_d \rightarrow \Lambda \psi$ can be used to constrain the parameter space of the dark antibaryon mass. BaBar set an upper limit of $\mathcal{B} < (0.13 - 5.2) \times 10^{-5}$ in the mass range $1.0 < m_{\psi} < 4.1\,\mathrm{GeV}$~\cite{BaBar:2023rer}, while Belle obtained $\mathcal{B} < 2.1 \times 10^{-5}$ for $1.0 < m_{\psi} < 3.9\,\mathrm{GeV}$~\cite{Belle:2021gmc}.

From Fig.~\ref{fig:Br}, one observes that the theoretical branching fraction for the $(\boldsymbol b)$ model is several orders of magnitude below the reported experimental limits across the entire allowed mass range. Consequently, no constraints can be derived on the dark antibaryon mass if the decay proceeds solely via the $\mathcal{O}^{(\boldsymbol b)}$ operator. This behavior is expected since the $(\boldsymbol b)$ model corresponds to a type-I interaction, for which the available phase space is more suppressed than in type-II and type-III interactions~\cite{Alonso-Alvarez:2021qfd}.

On the other hand, comparing the theoretical branching fraction for the $(\boldsymbol s)$-model with the BaBar exclusion limits reveals two allowed mass windows. The region $2.868\,\mathrm{GeV} < m_{\psi} < 3.650\,\mathrm{GeV}$ is excluded by the upper limit $\mathcal{B} < 5.2 \times 10^{-5}$:
\begin{equation}
    \begin{aligned}
        \text{Window I (Low Mass):}   & \quad 1.000~\mathrm{GeV} \le m_{\psi} \le 2.868~\mathrm{GeV}, \\
        \text{Window II (High Mass):} & \quad 3.650~\mathrm{GeV} \le m_{\psi} \le 4.164~\mathrm{GeV}.
    \end{aligned}
    \label{eq:MassWindows_BaBar}
\end{equation}
Since only the upper limits of the two couplings are available, the calculated branching fraction may be overestimated. Consequently,  the excluded mass interval could shrink, if the true coupling are smaller. For the stronger limit $\mathcal{B} < 0.13 \times 10^{-5}$, only the point $m_\psi = 4.164\,\mathrm{GeV}$ remains viable. Furthermore, comparing the theoretical branching fraction with the constraint from the Belle collaboration restricts the allowed mass region to
\begin{equation}
    4.112~\mathrm{GeV} \le m_{\psi} \le 4.164~\mathrm{GeV},
    \label{eq:MassWindows_Belle}
\end{equation}
which lies above the mass interval explored by Belle in the search for this decay channel.

%%%%%%%%%%%%%%%%%%%%%%%%%%%%%%%%%%%%%%%%%%%%%%%%%%%%%%%%%%%%%%%%%%%%
\section{Conclusion}
\label{sec:conclusion}
In the present work, we have investigated the exclusive decay $B_d\to\Lambda\psi$ within the framework of LCSR. Such processes arise naturally in the recently proposed $B$-mesogenesis scenario, in which $B$-meson decays can simultaneously address the observed BAU and the nature of dark matter. Furthermore, the branching fractions of these decay channels are potentially accessible at $B$ factories. A variety of inclusive and exclusive $B$-meson decay channels involving dark-sector particles has been studied in the literature. In particular, the decay mode $B_d\to\Lambda\psi$ provides a well-defined exclusive baryonic channel that can be directly confirmed in experimental searches. The upper limits on the branching fraction of this decay have been reported by the Belle and BaBar collaborations and are listed by the PDG.

To calculate the branching fraction of this decay, we employed the QCD light-cone sum rule approach and determined the $B_d\to\Lambda$ transition form factors using the $\Lambda$-baryon distribution amplitudes of Ref.~\cite{Liu:2008yg} up to twist-six accuracy. The resulting form factors for the two considered scenarios, namely the $(\boldsymbol s)$- and $(\boldsymbol b)$-models, were obtained in the region $-10\,\mathrm{GeV}^2 \le q^2 \le 0\,\mathrm{GeV}^2$ and subsequently extrapolated to the full kinematic domain using the $z$-expansion parametrization. Employing these form factors, we derived the decay widths and branching fractions for both models as functions of the dark antibaryon mass. By comparing the obtained branching fractions with the existing experimental upper limits reported by the Belle and BaBar collaborations, the allowed mass ranges for the dark antibaryon were determined.

We observe that two form factors in each model acquire nonzero contributions. 
Employing the couplings presented in Ref.~\cite{Lenz:2024rwi} the corresponding branching fractions restrict the kinematically allowed mass range of the dark-sector antibaryon. In particular, using the available upper limits of the couplings, the mass region $2.868\,\mathrm{GeV} < m_{\psi} < 3.650\,\mathrm{GeV}$ 
is excluded by the BaBar upper limit $\mathcal{B}(B_d\to\Lambda\psi) < 5.2\times10^{-5}$, while the more stringent Belle limit further constrains the allowed region to $4.112~\mathrm{GeV} \le m_{\psi} \le 4.164~\mathrm{GeV}$. Furthermore, our study shows that the twist-5 and -6 contributions for all four form factors of both models are suppressed compared to the leading-twist contributions, while the twist-4 contribution dominates over twist-3 in each form factor. The origin of this dominance, and its impact on the extracted form factors and branching fractions, and the dark antibaryon mass window require $ \Lambda $ DAs determined to higher accuracy, as well as including the higher-order QCD corrections to the sum rule.

 Our analysis demonstrates that this decay channel can serve as a sensitive probe of dark-sector $B$-mesogenesis scenarios. In particular, the decay chain $B_d\to \Lambda\psi \to p\pi\psi$ provides an experimentally favorable signature due to the reconstructible $\Lambda\to p\pi$ decay and the relatively stringent upper bounds on its branching fraction compared to several previously studied channels. A new study \cite{Hiller:2026osz} uses LHC missing energy searches for jets to derive updated constraints on the couplings relevant for $B$-mesogenesis scenario. The study finds that the couplings used in such scenarios involving light quark operators are more constrained than the values of \cite{Alonso-Alvarez:2021qfd,Lenz:2024rwi} adopted in our LCSR. Since the estimated branching ratio scales as $|G|^2$, these stronger bounds would reduce and may exclude $B_d \to \Lambda$ decay channel.

 In order to make a better comparison of the theoretical and experimental results, the $ \Lambda $ DAs and related parameters should be calculated with more accuracy and the experimental limits should be put more precisely. The higher order conformal spin (P-wave) corrections to the $\Lambda$ DAs have been obtained in Ref.~\cite{Liu:2014uha}. A study of these corrections along with the NLO QCD corrections mentioned in Subsec.~\ref{subsec:FFs}, is left for further works. Also, it would be interesting to see whether $B_d\to \Lambda$ form factors satisfy large-recoil symmetry relations similar to those established for $\Lambda_b \to \Lambda$ transitions \cite{Wang:2008sm}. This would require heavy quark effective theory/soft-collinear effective theory analysis of the three-quark operators in Eqs.~\eqref{eq:O_s} and~\eqref{eq:O_b}, left for future studies.
%%%%%%%%%%%%%%%%
\section*{ACKNOWLEDGMENTS} 
K. Azizi thanks Iran national science foundation (INSF) for the partial financial support supplied under the elites Grant No. 40405095. 
%
%%%%%%%%%%%%%%%%%%%%%%%%%%%%%%%%%%%%%%%%%%%%%%%%%%%%%%%%%%%%%%%%%%%%%%%
\onecolumngrid
\appendix
\section{$\Lambda$ Distribution Amplitudes}
\label{sec:appA}
The distribution amplitudes for $\Lambda$ are given with the conformal partial wave expansion approach:
\begin{equation}
    \begin{aligned}
         & 4 \langle 0| \epsilon^{abc} u^a_\alpha(a_1 x)\, d^b_\beta(a_2 x)\, s^c_\gamma(a_3 x) | \Lambda(p) \rangle                                                       \\
         & = \mathcal{S}_1\, m_\Lambda\, C_{\alpha\beta} (\gamma_5 \Lambda)_\gamma
        + \mathcal{S}_2\, m_\Lambda^2\, C_{\alpha\beta} (\slashed{x}\,\gamma_5 \Lambda)_\gamma  \\
         & + \mathcal{P}_1\, m_\Lambda (\gamma_5 C)_{\alpha\beta}\, \Lambda_\gamma
        + \mathcal{P}_2\, m_\Lambda^2 (\gamma_5 C)_{\alpha\beta} (\slashed{x}\,\Lambda)_\gamma  \\&
          + \left(\mathcal{V}_1 + \frac{x^2 m_\Lambda^2}{4}\, \mathcal{V}_1^M \right) (\slashed{p} C)_{\alpha\beta} (\gamma_5 \Lambda)_\gamma 
         + \mathcal{V}_2\, m_\Lambda (\slashed{p} C)_{\alpha\beta} (\slashed{x}\,\gamma_5 \Lambda)_\gamma 
          + \mathcal{V}_3\, m_\Lambda (\gamma_\mu C)_{\alpha\beta} (\gamma^\mu \gamma_5 \Lambda)_\gamma \\&
         + \mathcal{V}_4\, m^2_\Lambda (\slashed{x} C)_{\alpha\beta} (\gamma_5 \Lambda)_\gamma 
          + \mathcal{V}_5\, m_\Lambda^2 (\gamma_\mu C)_{\alpha\beta} (i\sigma^{\mu\nu} x_\nu \gamma_5 \Lambda)_\gamma
        + \mathcal{V}_6\, m_\Lambda^3 (\slashed{x} C)_{\alpha\beta} (\slashed{x}\,\gamma_5 \Lambda)_\gamma  \\
        & + \left(\mathcal{A}_1 + \frac{x^2 m_\Lambda^2}{4}\, \mathcal{A}_1^M \right) (\slashed{p}\,\gamma_5 C)_{\alpha\beta} \Lambda_\gamma 
         + \mathcal{A}_2\, m_\Lambda (\slashed{p}\,\gamma_5 C)_{\alpha\beta} (\slashed{x}\,\Lambda)_\gamma + \mathcal{A}_3\, m_\Lambda (\gamma_\mu \gamma_5 C)_{\alpha\beta} (\gamma^\mu \Lambda)_\gamma \\
       & + \mathcal{A}_4\, m^2_\Lambda (\slashed{x}\,\gamma_5 C)_{\alpha\beta} \Lambda_\gamma + \mathcal{A}_5\, m_\Lambda^2 (\gamma_\mu \gamma_5 C)_{\alpha\beta} (i\sigma^{\mu\nu} x_\nu \Lambda)_\gamma 
         + \mathcal{A}_6\, m_\Lambda^3 (\slashed{x}\,\gamma_5 C)_{\alpha\beta} (\slashed{x}\,\Lambda)_\gamma  \\
         &  + \left(\mathcal{T}_1 + \frac{x^2 m_\Lambda^2}{4}\, \mathcal{T}_1^M \right) (p^\nu i\sigma_{\mu\nu} C)_{\alpha\beta} (\gamma^\mu \gamma_5 \Lambda)_\gamma 
          + \mathcal{T}_2\, m_\Lambda (x^\mu p^\nu i\sigma_{\mu\nu} C)_{\alpha\beta} (\gamma_5 \Lambda)_\gamma\\&
        + \mathcal{T}_3\, m_\Lambda (\sigma_{\mu\nu} C)_{\alpha\beta} (\sigma^{\mu\nu} \gamma_5 \Lambda)_\gamma  
           + \mathcal{T}_4\, m_\Lambda (p^\nu\sigma_{\mu\nu} C)_{\alpha\beta} (\sigma^{\mu\rho} x_\rho\gamma_5 \Lambda)_\gamma + \mathcal{T}_5\, m_\Lambda^2 (i\sigma_{\mu\nu} x^\nu C)_{\alpha\beta} (\gamma^\mu \gamma_5 \Lambda)_\gamma \\
        & + \mathcal{T}_6\, m_\Lambda^2 (x^\mu p^\nu i\sigma_{\mu\nu} C)_{\alpha\beta} (\slashed{x}\,\gamma_5 \Lambda)_\gamma + \mathcal{T}_7\, m_\Lambda^2 (\sigma_{\mu\nu} C)_{\alpha\beta} (\sigma^{\mu\nu} \slashed{x}\,\gamma_5 \Lambda)_\gamma 
        + \mathcal{T}_8\, m_\Lambda^3 (x^\nu\sigma_{\mu\nu} C)_{\alpha\beta} (\sigma^{\mu\rho} x_\rho\,\gamma_5 \Lambda)_\gamma,
    \end{aligned}
    \label{eq:DAs}
\end{equation}
where $\Lambda$ is the spinor and $\sigma_{\mu\nu}=\frac i2[\gamma_\mu,\gamma_\nu]$. The 24 calligraphic functions may not have a definite twist, but they can be written as linear combination of different DAs with different powers of the scalar product $p\cdot x$ . The explicit relations for the scalar, pseudo-scalar, vector, axial-vector and tensor DAs, $F=S_i,P_i,V_i,A_i,T_i$, with the calligraphic functions are as follows:
\begin{equation}
\begin{aligned}
&\mathcal{S}_1=S_1,&\qquad\qquad\qquad&\mathcal{S}_2 = \frac{1}{2p \cdot x}\left(S_1-S_2\right),\\
&\mathcal{P}_1=P_1,&\qquad\qquad\qquad &\mathcal{P}_2 = \frac{1}{2p \cdot x}\left(P_2-P_1\right),\\
&\mathcal{V}_1=V_1,&\qquad\qquad\qquad&\mathcal{V}_2 = \frac{1}{2p \cdot x}\left(V_1-V_2-V_3\right),\\
&\mathcal{V}_3= \frac{V_3}2,&\qquad\qquad\qquad&\mathcal{V}_4=\frac{1}{4 p \cdot x }\left(-2V_1+V_3+V_4+2V_5\right),\\
&\mathcal{V}_5 = \frac{1}{4 p \cdot x }\left(V_4-V_3\right),&\qquad\qquad\qquad&\mathcal{V}_6=\frac{1}{4(p \cdot x)^2}\left(-V_1+V_2+V_3+V_4+V_5-V_6\right),\\
&\mathcal {V}_1^M(x_2)=\int_0^{1-x_2} dx_1V_1^M(x_1,x_2,1-x_1-x_2),\\
&\mathcal{A}_1=A_1,&\qquad\qquad\qquad&\mathcal{A}_2 = \frac{1}{2p \cdot x }\left(-A_1+A_2-A_3\right),\\
&\mathcal{A}_3 = \frac{A_3}2,&\qquad\qquad\qquad&\mathcal{A}_4=\frac{1}{4 p \cdot x }\left(-2A_1-A_3-A_4+2A_5\right),\\
&\mathcal{A}_5=\frac{1}{4 p \cdot x }\left(A_3-A_4\right),&\qquad\qquad\qquad&\mathcal{A}_6=\frac{1}{4(p \cdot x)^2 }\left(A_1-A_2+A_3+A_4-A_5+A_6\right),\\
&\mathcal A_1^M(x_2)=\int_0^{1-x_2} dx_1A_1^M(x_1,x_2,1-x_1-x_2),\\
&\mathcal{T}_1=T_1,&\qquad\qquad\qquad &\mathcal{T}_2 =\frac{1}{2p \cdot x}\left(T_1+T_2-2T_3\right),\\
&\mathcal{T}_3= \frac{T_7}2,&\qquad\qquad\qquad&\mathcal{T}_4=\frac{1}{2 p \cdot x }\left(T_1-T_2-2T_7\right),\\
&\mathcal{T}_5=\frac{1}{ 2 p \cdot x }\left(-T_1+T_5+2T_8\right),&\qquad\qquad\qquad&\mathcal{T}_6=\frac{1}{4(p \cdot x)^2 }\left(2T_2-2T_3-2T_4+2T_5+2T_7+2T_8\right),\\
&\mathcal{T}_7=\frac{1}{4 p \cdot x }\left( T_7-T_8\right),&\qquad\qquad\qquad&\mathcal{T}_8=\frac{1}{4(p \cdot x)^2}\left(-T_1+T_2+T_5-T_6+2T_7+2T_8\right),\\
&\mathcal T_1^M(x)=V_1^M(x)+A_1^M(x).
\end{aligned}
\end{equation}

 Each distribution amplitude $F(a_ip\cdot x)$ can be expressed in terms of twist amplitudes as:
\begin{equation}
    F(a_ip\cdot x)=\int dx_1\,dx_2\,dx_3 \,\delta(x_1+x_2+x_3-1) \exp{-ip\cdot x \sum_i x_i a_i}F(x_i),
    \label{eq:QCD:DA:generic}
\end{equation}
where $x_i$ with $i=1,2,3$ are longitudinal momentum fractions carried by valance quarks of $\Lambda$-baryon. The explicit representations of the distribution amplitudes $F$ with definite twists are presented in Refs.~\cite{Liu:2008yg,Aliev:2010uy,Liu:2009mb,Liu:2009sn}.
%%%%%%%%%%%%%%%%%%%%%%

\section{Borel Transformation and Continuum Subtraction}
\label{sec:appB}
The Borel transformation and continuum subtractions are performed using the replacement rules bellow \cite{Dehghan:2025ncw,Braun:2006hz}:
 \begin{equation}
\begin{aligned}
\int dx\frac{\varrho(x)}{(q-xP)^2}& \to -\int_{x_0}^1\frac{dx}{x}\varrho(x)e^{-s(x)/M^2}, \\
\int dx\frac{\varrho(x)}{(q-xP)^4}& \to \frac{1}{M^2}\int_{x_0}^1\frac{dx}{x^2}\varrho(x)e^{-s(x)/M^2} 
 + \frac{\varrho(x_0)}{Q^2+x_0^2 m_\Lambda^2}e^{-s_0/M^2}, \\
\int dx\frac{\varrho(x)}{(q-xP)^6}& \to -\frac{1}{2M^4}\int_{x_0}^1\frac{dx}{x^3}\varrho(x)e^{-s(x)/M^2} 
 -\frac{1}{2 M^2}\frac{\varrho(x_0)}
               {x_0(Q^2+x_0^2 m_\Lambda^2)}e^{-s_0/M^2}
  + \frac{1}{2}\frac{x_0^2}{Q^2+x_0^2m_\Lambda^2}
    \\&\times\left[\frac{d}{dx_0}
     \left(\frac{\varrho(x_0)}
     {x_0(Q^2+x_0^2 m_\Lambda^2)}\right)\right]e^{-s_0/M^2},\\[6pt]
 \int dx\frac{P'^2\varrho(x)}{(q-xP)^2}&\to-\int_{x_0}^1\frac{dx}{x}\varrho(x)s(x)e^{-s(x)/M^2},\\
 \int dx\frac{P'^2\varrho(x)}{(q-xP)^4}& \to \int_{x_0}^1\frac{dx}{x^2}\varrho(x)\left(-1+\frac{s(x)}{M^2}\right)e^{-s(x)/M^2} 
 + \frac{\varrho(x_0)}{Q^2+x_0^2 m_\Lambda^2}s_0e^{-s_0/M^2}, \\
  \int dx\frac{P'^2\varrho(x)}{(q-xP)^6}& \to \frac{1}{M^2}\int_{x_0}^1\frac{dx}{x^3}\varrho(x)\left(1-\frac{s(x)}{2M^2}\right)e^{-s(x)/M^2}  +\frac{1}{2}\frac{\varrho(x_0)}
               {x_0(Q^2+x_0^2 m_\Lambda^2)}\left(1-\frac{s_0}{M^2}\right)e^{-s_0/M^2}
  \\&+ \frac{1}{2}\frac{x_0^2}{Q^2+x_0^2m_\Lambda^2}
    \left[\frac{d}{dx_0}
     \left(\frac{\varrho(x_0)}
     {x_0(Q^2+x_0^2 m_\Lambda^2)}\right)\right]s_0e^{-s_0/M^2},\\[6pt]
   \int dx\frac{P'^4\varrho(x)}{(q-xP)^4}& \to \int_{x_0}^1\frac{dx}{x^2}\varrho(x)\left(-2s(x)+\frac{s^2(x)}{M^2}\right)e^{-s(x)/M^2} 
 + \frac{\varrho(x_0)}{Q^2+x_0^2 m_\Lambda^2}s^2_0e^{-s_0/M^2}, \\
 \int dx\frac{P'^4\varrho(x)}{(q-xP)^6}& \to \int_{x_0}^1\frac{dx}{x^3}\varrho(x)\left(-1+\frac{2s(x)}{M^2}-\frac{s^2(x)}{2M^4}\right)e^{-s(x)/M^2}  +\frac{\varrho(x_0)}
               {x_0(Q^2+x_0^2 m_\Lambda^2)}\left(s_0-\frac{s^2_0}{2M^2}\right)e^{-s_0/M^2}
  \\&+ \frac{1}{2}\frac{x_0^2}{Q^2+x_0^2m_\Lambda^2}
    \left[\frac{d}{dx_0}
     \left(\frac{\varrho(x_0)}
     {x_0(Q^2+x_0^2 m_\Lambda^2)}\right)\right]s^2_0e^{-s_0/M^2},
\end{aligned}
\label{eq:Borel}
\end{equation}
where $P^{\prime\,2} = (P + q)^2$, $M^2$ denotes the Borel parameter, and $s(x)$ is defined as:
\begin{equation}
    s(x) = \frac1x\biggl(m_b^2 + (1-x)Q^2 + (1-x)x\, m_\Lambda^2\biggr).
    \label{eq:s_definition}
\end{equation}

Here the momentum transfer is defined as $Q^2 \equiv -q^2$. The parameter $x_0$ is obtained from the condition that Eq.~\eqref{eq:s_definition} reaches the continuum threshold $s=s_0$:
\begin{equation}
          x_0  =\frac1{2m_\Lambda^2}\biggl[\sqrt{(Q^2 + s_0 - m_\Lambda^2)^2 + 4m_\Lambda^2(Q^2 + m_b^2)}-\left(Q^2 + s_0 - m_\Lambda^2\right)\biggr].
    \label{eq:x0_definition}
\end{equation}

The variables $s$ and $x_0$ must be modified accordingly for terms involving partial integrations arising from the elimination of inverse powers of $P\cdot x$. For instance, if a term contains two integrals $\int_0^1 dx_i\int_0^{x_i}d\alpha$ instead of a single one with respect to $x_i$, the variable $s$ becomes a function of $\alpha$, and an analogous parameter $\alpha_0$ must be defined in place of $x_0$.
%%%%%%%%%%%%%%%%%%%%%%%%%%%%%%%%%%
\section{Explicit Expression for $\mathcal B_{M^2}\Pi^{(\boldsymbol s),(\boldsymbol b)}_{R}$ and $\mathcal B_{M^2}\tilde\Pi^{(\boldsymbol s),(\boldsymbol b)}_{L}$}
\label{sec:appC}
There are two functions with nonvanishing QCD side in model $(\boldsymbol s)$, namely the left-hand sides of Eqs.~\eqref{eq:FF_Borel}. The explicit expression for  $\mathcal B_{M^2}\Pi^{(\boldsymbol s)}_R$ reads:
\begin{equation}
\label{eq:ExplicitPiRs}
\begin{aligned}
\mathcal B_{M^2}\Pi^{(\boldsymbol s)}_R(Q^2,M^2,s_0)&=-\frac{m_\Lambda^4}{2 M^4}\int_{x_0}^1 d\bar\beta\int_{\bar\beta}^1 d\bar\alpha\int_{\bar\alpha}^1 dx_2\int_0^{1-x_2} dx_1e^{-s(\bar\beta)/M^2}\left(A_{123456} + V_{123456}\right)\Big[I_4(\beta)+ Q^2\Big]
\\&+\int_{x_0}^{1} d\bar\alpha\int_{\bar\alpha}^{1} dx_2\int_{0}^{1-x_2} dx_1
\Bigg\{e^{-s(\bar\alpha)/M^2} \frac{m_\Lambda^2}{4\bar\alpha M^2}\Bigg(\left(A_{123}+V_{123}\right) \Big[I_4(\alpha)+\bar\alpha^2 m_\Lambda^2\Big]
\\&+\left(A_{1345}-V_{1345}\right)\Big[I_4(\alpha)+Q^2\Big]+2\bar\alpha m_\Lambda m_b\left(P_{21}-S_{12}\right)\Bigg)
+e^{-s_0/M^2}\frac{m_\Lambda^4 x_0^2}{2 M^2 (I_1+m_\Lambda^2 x_0^2)^3}
\\&\times\left(A_{123456} + V_{123456}\right)\Bigg(-I_1^2 Q^2+ x_0I_1\Big[I_1\left(I_2(x_0)+4M^2\right)-2m_b^2M^2\Big]
- x_0^2I_1\Big[m_\Lambda^2 \left(m_b^2+Q^2\right)
\\&-3M^2\left(Q^2+s_0+m_\Lambda^2\right)\Big]+2 x_0^3m_\Lambda^2 I_1I_2(x_0)
+ x_0^4m_\Lambda^2\Big[M^2\left(Q^2+s_0\right)+ m_\Lambda^2\left(M^2+Q^2-2m_b^2\right)\Big]
\\&+ x_0^5 m_\Lambda^4I_2(x_0)-x_0^6m_\Lambda^6\Bigg)\Bigg\}
\\&-\int_{x_0}^{1} dx_2\int_{0}^{1-x_2} dx_1\Bigg\{\frac{1}{4 x_2}e^{-s(x_2)/M^{2}}\Bigg(2 m_\Lambda m_b \left(P_{1}+S_1\right)+m_\Lambda^{2} x_2 \left(A_{3}+V_3\right)+\left(V_1-A_1\right)I_3(x_2)\Bigg)
\\&+ e^{-s0/M^2}\frac{m_\Lambda^2x_0}{4(I_1+m_\Lambda^2x_0^2)}\Bigg(2m_\Lambda m_b x_0\left(S_{12}-P_{21}\right)+\left(A_{123}+V_{123}\right)x_0\Big[I_3(x_0)-m_\Lambda^2(1-x_0)\Big]
\\&+\left(V_{1345}-A_{1345}\right)\Big[Q^2-x_0(I_3(x_0)+m_\Lambda^2x_0)\Big]\Bigg)\Bigg\}
\\&+\int_{x_0}^1dx_2 e^{-s(x_2)/M^2}\frac{m_\Lambda^2}{4M^2x_2}\big[V_1^M(x_2)-A_1^M(x_2)\big]\Bigg(M^2x_2\bigg[2Q^2(x_2+1)-M^2\bigg]-I_3(x_2)I_5(\bar x_2)\Bigg)
\\&+e^{s_0/M^2}\frac{m_\Lambda^2x_0}{4M^2(I_1+m_\Lambda^2x_0^2)^3}
\big[V_1^M(x_0)-A_1^M(x_0)\big]\Bigg\{\left(I_1+m_\Lambda^2x_0^2\right)^2I_3(x_0)\Big[m_\Lambda^2 x_0 - (1 + x_0)
\\&\times (Q^2 + s_0 x_0)\Big]+M^2I_1^2\Bigg(I_3(x_0)(-1+3x_0+x_0^2)+m^2(1-2x_04x_0^3)-s_0(1+x_0)^2\Bigg)
\\&+M^2m_\Lambda^2x_0^3\Bigg(I_3^2(x_0)+I_3(x_0)(1+x_0)\Big[s_0(1-x_0)+m_\Lambda^2(-1+4x_0+x_0^2)\Big]+m_\Lambda^2x_0\Big[s_0(1+x_0)^2
\\&+m_\Lambda^2(-1-2x_0+4x_0^2+4x_0^3)\Big]\Bigg)+M^2I_1x_0\Bigg(-I_3^2(x_0)(1+2x_0)-I_3(x_0)s_0(1+x_0)(1+3x_0)
\\&+4m_\Lambda^4x_0^2(-1x_0+2x_0^2)+I_3(x_0)m_\Lambda^2\Big[2x_0(1+x_0)^2+1\Big]\Bigg)\Bigg\},
\end{aligned}
\end{equation}
and $\mathcal B_{M^2}\tilde\Pi^{(\boldsymbol s)}_L$ is as follows:
\begin{equation}
\label{eq:ExplicitTildePiLs}
\begin{aligned}
\mathcal B_{M^2}\tilde\Pi^{(\boldsymbol s)}_L(Q^2,M^2,s_0)&=-\int_{x_0}^1 d\bar\beta\int_{\bar\beta}^1 d\bar\alpha\int_{\bar\alpha}^1 dx_2\int_0^{1-x_2} dx_1\frac{m_\Lambda^3}{2 \bar\beta M^4}e^{-s(\bar\beta)/M^2}\left(A_{123456} + V_{123456}\right)\Big[I_4(\beta)+ Q^2\Big]
\\&+\int_{x_0}^{1} d\bar\alpha\int_{\bar\alpha}^{1} dx_2\int_{0}^{1-x_2} dx_1
\Bigg\{e^{-s(\bar\alpha)/M^2} \frac{m_\Lambda}{4\bar\alpha M^2}\Bigg(-I_3(\bar\alpha)(A_{123}+V_{123})+2m_\Lambda m_b(P_{21}-S_{12})\Bigg)
\\&+e^{-s_0/M^2}\frac{m_\Lambda^3 x_0}{2 M^2 (I_1+m_\Lambda^2 x_0^2)^3}\left(A_{123456} + V_{123456}\right)\Bigg(-I_1^2 Q^2+ x_0I_1\Big[I_1\left(I_2(x_0)+3M^2\right)-m_b^2M^2\Big]
\\&- x_0^2I_1\Big[m_\Lambda^2 \left(m_b^2+Q^2\right)-2M^2\left(Q^2+s_0+m_\Lambda^2\right)\Big]+ x_0^3m_\Lambda^2 (2I_1I_2(x_0)+m_b^2M^2)
- x_0^4m_\Lambda^4(I_1+m_b^2)
\\&+ x_0^5 m_\Lambda^4(I_2(x_0)+M^2)-x_0^6m_\Lambda^6\Bigg)\Bigg\}
\\&-\int_{x_0}^{1} dx_2\int_{0}^{1-x_2} dx_1\Bigg(\frac{m_\Lambda}{4 x_2}e^{-s(x_2)/M^{2}}\left(A_{3}+V_3\right)+e^{-s_0/M^2}\frac{m_\Lambda x_0}{4(I_1+m_\Lambda^2x_0^2)}\Bigg[I_3(x_0)(A_{123}+V_{123})
\\&-2m_bm_\Lambda(P_{21}-S_{12})\Bigg]\Bigg).
\end{aligned}
\end{equation}
 The corresponding correlation functions for model $(\boldsymbol b)$ enter the LCSR calculation in an analogous way. The correlator $\mathcal B_{M^2}\Pi_R^{(\boldsymbol b)}$ is given by
\begin{equation}
\label{eq:ExplicitPiRb}
\begin{aligned}
\mathcal B_{M^2}\Pi_R^{(\boldsymbol b)}(Q^2,M^2,s_0)&=\int_{x_0}^1 d\bar\beta\int_{\bar\beta}^1 d\bar\alpha\int_{\bar\alpha}^1 dx_2\int_0^{1-x_2} dx_1e^{-s(\bar\beta)/M^{2}}\frac{m_{\Lambda}^{3}}{4\bar\beta M^{4}}\Bigg(I_5(\beta)\Big[\bar\beta m_{\Lambda}(A_{123456}+V_{123456})
\\&+m_{b}T_{234578}\Big]-3m_{b}\big[I_5(\beta)+\bar\beta M^2\big]T_{125678}\Bigg)
\\&+\int_{x_0}^{1} d\bar\alpha\int_{\bar\alpha}^{1} dx_2\int_{0}^{1-x_2} dx_1
\Bigg\{e^{-s(\bar\alpha)/M^2}\frac{m_\Lambda}{8\bar\alpha^2 M^2}\Bigg(m_\Lambda\bar\alpha(I_4(\alpha)+Q^2)\Big[A_{1345}-V_{1345}
\\&+3(V_{43}-A_{34})\Big]+\bar\alpha^2 m_\Lambda^2 m_b\Big[2(P_{21}-S_{12})+6(T_{158}+2T_{78})-T_{234578}\Big]+2m_\Lambda\bar\alpha\Big[\bar\alpha  (\bar\alpha m_\Lambda^2 + M^2) 
\\&- 2 m_\Lambda Q^2\Big](A_{123}+V_{123})+m_b\bar\alpha\Big[I_2(\alpha)+5M^2\Big]T_{123}-4m_bI_4(\alpha) T_{127}\Bigg)
+e^{-s_0/M^2}
\\&\times\frac{3m_\Lambda^3 x_0m_b}{4M^2(I_1+x_0^2m_\Lambda^2)^3}\bigg[T_{125678}+T_{234578}+A_{123456}+V_{123456}\bigg]\Bigg((I_1+m_\Lambda^2x_0^2)^2 \bigg[-I_5(\bar x_0)
\\&+M^2(1+2x_0(x_0-1))\bigg]+Q^2(3I_1-m_\Lambda^2x_0^2)-x_0^2s_0(I_1+m_\Lambda^2x_0^2)-2I_1(m_\Lambda^2+s_0)\Bigg)\Bigg\}
\\&-\int_{x_0}^{1} dx_2\int_{0}^{1-x_2} dx_1\Bigg\{e^{-s(x_2)/M^2}\frac{m_\Lambda}{8x_2}\Bigg(m_b\bigg[2(P_1+S_1)+6(T_1-2T_7)-T_{127}\bigg]+2m_\Lambda x_2\bigg[A_1-V_1
\\&+2(A_3+V_3)\bigg]\Bigg)+e^{-s_0/M^2}\frac{m_\Lambda x_0}{8(I_1+m_\Lambda^2x_0^2)}\Bigg(m_\Lambda^2m_b x_0\bigg[2(P_{21}-S_{12})+6(T_{158}+2T_{127})-T_{234578}\bigg]
\\&+m_b\bigg[4(I_3(x_0)+m_\Lambda^2x_0)T_{127}+(I_3(x_0)+2m_\Lambda^2x_0)T_{123}\bigg]-2m_\Lambda(Q^2-m_\Lambda^2x_0^2)(A_{123}+V_{123})
\\&+m_\Lambda\bigg(Q^2-x_0(I_3(x_0)+m_\Lambda^2x_0)\bigg)\bigg[A_{1345}-V_{1345}+3(V_{43}-A_{34})\bigg]\Bigg)\Bigg\}
\\&+\int_{x_0}^1dx_2 e^{-s(x_2)/M^2}\frac{m^2_\Lambda}{4M^4 x_2}\Bigg(\bigg[M^2(I_4(\bar x_2)+Q^2)-m_\Lambda^2 x_2I_5(\bar x_2)\bigg]\bigg[A_1^M(x_2)-V_1^M(x_2)\bigg]
\\&-3m_\Lambda m_b I_5(\bar x_2)T_1^M(x_2)\Bigg)+e^{-s_0/M^2}\frac{m^2_\Lambda x_0}{4 M^2 (I_1 + m_\Lambda^2 x_0^2)^3}\Bigg\{\bigg[A_1^M(x0)-V_1^M(x_0)\bigg]\Bigg( (I_1+m_\Lambda^2 x_0^2)^2
\\&\times\bigg[M^2\bigg(m_\Lambda^2 x_0^2(x_0+2)-2s_0 x_0^2+Q^2(1-2x_0)\bigg)-m_\Lambda^2x_0 I_5(\bar x_0)\bigg]+M^2 x_0 I_1^2\bigg[Q^2-m_\Lambda^2
\\&+s_0(1+2x_0)\bigg]+I_1 x_0^2 M^2 m_\Lambda^2\bigg[m_\Lambda^2 x_0-s_0 x_0-Q^2(x_0+2)\bigg]\Bigg)+3m_\Lambda m_bT_1^M(x_0)\Bigg((I_1+m_\Lambda^2 x_0^2)^2
\\&\times\bigg[M^2 x_0(1+x_0)-I_5(\bar x_0)\bigg]+m^2 m_\Lambda^2 x_0^3(Q^2-s_0 x_0^2)-I_1 M^2 x_0\bigg[Q^2(1+2x_0)-2m_\Lambda^2 x_0
\\&+s_0 x_0(2+3x_0)\bigg]\Bigg)\Bigg\},
\end{aligned}
\end{equation}
and for $\mathcal B_{M^2}\tilde\Pi_L^{(\boldsymbol b)}$ we obtain
\begin{equation}
\label{eq:ExplicitTildePiLb}
\begin{aligned}
\mathcal B_{M^2}\tilde\Pi_L^{(\boldsymbol b)}(Q^2,M^2,s_0)&=\int_{x_0}^1 d\bar\beta\int_{\bar\beta}^1 d\bar\alpha\int_{\bar\alpha}^1 dx_2\int_0^{1-x_2} dx_1e^{-s(\bar\beta)/M^{2}}\frac{m_{\Lambda}^{2}}{4\bar\beta M^{2}}\Bigg(I_5(\beta)m_{\Lambda}\bigg[A_{123456}+V_{123456}\bigg]
\\&-m_{b}M^2T_{234578}\Bigg)
\\&+\int_{x_0}^{1} d\bar\alpha\int_{\bar\alpha}^{1} dx_2\int_{0}^{1-x_2} dx_1
\Bigg\{e^{-s(\bar\alpha)/M^2}\frac{m_\Lambda}{8\bar\alpha M^2}\Bigg(2\big(M^2-I_3(\bar\alpha)\big)\big[A_{123}+V_{123}\big]+m_\Lambda m_b\bigg[2(P_{21}
\\&-S_{12}+T_{123}+2T_{127})+6(T_{158}+2T_{78})-T_{234578}\bigg]\Bigg)+e^{-s_0/M^2}\frac{m_\Lambda^2 x_0}{4M^2(I_1+m_\Lambda^2 x_0^2)^3}
\\&\times\Bigg(m_\Lambda\bigg[A_{123456}+V_{123456}\bigg]\bigg((I_1+m_\Lambda^2 x_0^2)^2\Big[I_5(\bar x_0)-M^2 x_0(1+x_0)\Big]+m_\Lambda^2M^2 x_0^2(Q^2+s_0 x_0^2)
\\&-M^2 x_0 I_1\Big[Q^2(1+2x_0)-2x_0m_\Lambda^2-s_0 x_0(2+3x_0)\Big]\bigg)-m_b M^2(I_1+m_\Lambda^2 x_0^2)^2 T_{234578}\Bigg)\Bigg\}
\\&-\int_{x_0}^{1} dx_2\int_{0}^{1-x_2} dx_1\Bigg(\frac{m_\Lambda}{4 x_2}e^{-s(x_2)/M^{2}}\bigg[A_{1}-V_1+2(A_{3}-V_3)\bigg]+e^{-s_0/M^2}\frac{m_\Lambda x_0}{8(I_1+m_\Lambda^2x_0^2)}
\\&\times\Bigg[2I_3(x_0)\bigg(A_{123}+V_{123}\bigg)
-m_bm_\Lambda \bigg(2 [P_{21} -  S_{12}+T_{123}+ 2 T_{127}]+ 6[ T_{158}+2T_{78}] - T_{234578}\bigg)\Bigg]\Bigg)
\\&+\int_{x_0}^1dx_2 e^{-s(x_2)/M^2}\frac{m^2_\Lambda}{4M^4 x_2}\Bigg(mI_5(\bar x_2)\bigg[V_1^M(x_2)-A_1^M(x_2)\bigg]+3m_b M^2T_1^M(x_2)\Bigg)
\\&+e^{-s_0/M^2}\frac{m_\Lambda^2 x_0}{4 M^2 (I_1 + m_\Lambda^2 x_0^2)^3}\Bigg(\bigg[A_1^M(x_0)-V_1^M(x_0)\bigg]\bigg(m(I_1+m_\Lambda^2 x_0^2)^2\Big[M^2 x_0(1+x_0)-I_5(\bar x_0)\Big]
\\&+m_\Lambda M^2 x_0\Big[(Q^2-s_0 x_0^2)(m_\Lambda^2 x_0-I_1(1+2x_0))-2s_0 x_0^2 I_1\Big]\bigg)+3 m_b M^2 (I_1 + m_\Lambda^2 x_0^2)^2T_1^M(x_0)\Bigg),
\end{aligned}
\end{equation}
where $\bar x=1-x$.  The definitions of $I_i$ are as follows

\begin{equation}
\begin{aligned}
&I_1=m_b^2-Q^2,&\qquad\qquad &I_2(x)=m_\Lambda^2-4M^2+Q^2-s(x),\\
&I_3(x)=m_\Lambda^2(1-2x)+Q^2-s(x),&\qquad\qquad &I_4(x)=\bar x\big[M^2-Q^2-xm_\Lambda^2+s(\bar x)\big],\\
&I_5(x)=\bar x\big[m_\Lambda^2-(1+\bar x)s(\bar x)-(1-2\bar x)M^2\big]-(1+\bar x)Q^2.
\end{aligned}
\end{equation}

Here, we employ the following abbreviations from \cite{Braun:2000kw, Braun:2001tj, Braun:2006hz, Khodjamirian:2011jp, Boushmelev:2023huu}:
\begin{equation}
\begin{aligned}
&S_{12}=S_1-S_2,& \qquad\qquad\qquad &P_{21}=P_2-P_1,\\
&V_{43}=V_4-V_3,&\qquad\qquad\qquad &A_{34}=A_3-A_4,\\
&T_{78}=T_7-T_8,&\qquad\qquad\qquad &A_{123}=-A_1+A_2-A_3,\\
&V_{123}=V_1-V_2-V_3,&\qquad\qquad\qquad &T_{127}=T_1-T_2-2T_7,\\
&T_{123}=T_1+T_2-2T_3,&\qquad\qquad\qquad&T_{158}=-T_1+T_5+2T_8 ,\\
&V_{1345}=-2V_1 + V_3 + V_4 + 2V_5 ,& \qquad\qquad\qquad &A_{1345}=-2A_1-A_3-A_4+2A_5, \\
&V_{123456}=-V_1 + V_2 + V_3 + V_4 + V_5-V_6 ,&\qquad\qquad\qquad & A_{123456}=A_1-A_2 + A_3 + A_4-A_5 + A_6,\\
&T_{234578}=2T_2-2T_3-2T_4+2T_5+2T_7+2T_8,& \qquad\qquad\qquad &T_{125678}=-T_1 + T_2 + T_5-T_6 + 2T_7 + 2T_8,
\end{aligned}
\end{equation}
where for all the functions used above, $F_i=F_i(x_1,x_2,1-x_1-x_2)$.
\twocolumngrid

\bibliographystyle{apsrev4-1}
\bibliography{articlebiblio}
 
\end{document}